\documentclass[12pt]{article}

\usepackage{authblk}
\usepackage{graphicx}
\usepackage{amssymb}
\usepackage{tabularx,ragged2e,booktabs,caption}
\usepackage{multirow}
\usepackage{caption}
\usepackage{subcaption}
\usepackage{arydshln}
\usepackage[left, modulo]{lineno}
\usepackage{xurl}
\usepackage{amsmath}
\usepackage{hyperref}
\usepackage[misc]{ifsym}
\usepackage[superscript, nomove]{cite}
\usepackage{fullpage}
\usepackage{enumitem}
\usepackage{setspace}
\usepackage{textcomp}

\date{}

\newcommand{\aj}{Astron. J.}   % Astronomical Journal
\newcommand\ion[2]{#1$~${\scshape{#2}}}

\newcommand{\kms}{km~s$^{-1}$}

\newcommand{\HI}{\ion{H}{i}}

\newcommand{\CII}{\ion{C}{ii}}

\newcommand{\SiII}{\ion{Si}{ii}}
\newcommand{\SiIII}{\ion{Si}{iii}}

\title{\bfseries Diverse Metallicities of Fermi Bubble Clouds Indicate Dual Origins in the Disk and Halo}

\author{Trisha Ashley$^{1}$\href{mailto:tashley@stsci.edu}{\Letter}, Andrew J. Fox$^2$. Frances H. Cashman$^1$, Felix J. Lockman$^3$, Rongmon Bordoloi$^4$, Edward B. Jenkins$^5$, Bart P. Wakker$^6$, Tanveer Karim$^7$}

\begin{document}

%TC:ignore

\maketitle

\begin{enumerate}[label={\upshape$^{\arabic{enumi}}$}]
\setlength{\itemsep}{0pt}
\item{Space Telescope Science Institute, 3700 San Martin Drive, Baltimore, MD 21218}
\item{AURA for ESA, Space Telescope Science Institute, 3700 San Martin Drive, Baltimore, MD 21218}
\item{Green Bank Observatory, P.O. Box 2, Rt. 28/92, Green Bank, WV 24944, USA}
\item{Department of Physics, North Carolina State University, 421 Riddick Hall, Raleigh, NC 27695-8202}
\item{Department of Astrophysical Sciences, Princeton University, Princeton, NJ 08544, USA}
\item{Department of Astronomy, University of Wisconsin-Madison, 475 North Charter Street, Madison, WI 53706, USA}
\item{Center for Astrophysics $|$ Harvard and Smithsonian, 60 Garden Street, Cambridge, MA 02138, USA}
\end{enumerate}

\clearpage
\begin{abstract}

The Galactic Center is surrounded by two giant plasma lobes known as the Fermi Bubbles, extending $\sim$10 kpc both above and below the Galactic plane.  Spectroscopic observations of Fermi Bubble directions at radio, ultraviolet, and optical wavelengths have detected multi-phase gas clouds thought to be embedded within the bubbles referred to as Fermi Bubble high-velocity clouds (FB HVCs). While these clouds have kinematics that can be modeled by a biconical nuclear wind launched from the Galactic center, their exact origin is unknown because, until now, there has been little information on their heavy-metal abundance (metallicity). Here we show that FB HVCs have a wide range of metallicities from $<$20\% solar to $\sim$320\%\ solar. This result is based on the first metallicity survey of FB HVCs. These metallicities challenge the previously accepted tenet that all FB HVCs are launched from the Galactic center into the Fermi Bubbles with solar or super-solar metallicities. Instead, we suggest that FB HVCs originate in both the Milky Way's disk and halo. As such, some of these clouds may characterize circumgalactic medium that the Fermi Bubbles expand into, rather than material carried outward by the nuclear wind, changing the canonical picture of FB HVCs. More broadly, these results reveal that nuclear outflows from spiral galaxies can operate by sweeping up gas in their halos while simultaneously removing gas from their disks.

\end{abstract}

\section*{Main}

Nuclear outflows are an important source of feedback in spiral galaxies. In the Milky Way, evidence of a nuclear outflow is provided by two large plasma bubbles launched from the Galactic center into the halo, known as the Fermi Bubbles. Simulations and observations suggest that the Fermi Bubbles are likely the result of Galactic nuclear activity from Sagittarius A*$^{\mathrm[}$\cite{Guo_2012, Mou_2014, Miller_2016b, predehl2020}$^{\mathrm]}$ and that they possibly formed during a Seyfert flare event $\sim$3.5 Myr ago that ionized the Magellanic Stream\cite{Bland_Hawthorn_2013, Bland_Hawthorn_2019, Fox_2020}. However, star formation in the Galactic center cannot be ruled out as a partial contributor to the origin and growth of the bubbles\cite{Crocker_2015, Sarkar_2015}. Their large angular extent ($\pm$55\textdegree\ in latitude and $\pm$20\textdegree\ in longitude\cite{Su_2010, Dobler_2010, Ackermann_2014}) allows us to resolve a nuclear outflow and study its effect on the baryons of a large spiral galaxy in unparalleled detail. 

The bubbles have been detected across the electromagnetic spectrum in gamma-rays, microwaves, polarized radio emission, and X-rays \cite{Bland_Hawthorn_2003, Dobler_2008, Carretti_2013, predehl2020}. Spectroscopic studies in UV
metal line absorption, and atomic hydrogen and molecular emission have revealed populations of cool gas clouds embedded within the Bubbles\cite{Fox_2015, Bordoloi_2017, Savage_2017, Karim_2018, Ashley_2020, McClure_Griffiths_2013, Di_Teodoro_2018, Lockman_2020, Di_Teodoro_2020}. These Fermi Bubble high-velocity clouds (FB HVCs) are identified by their projected location towards the gamma-ray-defined bubbles and their velocities, which cannot be accounted for by Galactic rotation (typically exceeding absolute velocities of $\sim$90 km~s$^{-1}$). FB HVCs are thought to be associated with the bubbles because their UV-absorption covering fraction (80\%) far exceeds that of HVCs directly outside of the Fermi Bubbles  that are similarly not co-rotating with the Milky Way disk (28\%)$^{[}$\cite{Bordoloi_2017}$^{]}$.

Previously, all FB HVCs have been assumed to originate in the disk of the Milky Way, becoming entrained in the Fermi Bubbles and traveling to higher Galactic latitudes\cite{Fox_2015, Bordoloi_2017, Ashley_2020}. FB HVCs detected in H\;{\small{\MakeUppercase{\romannumeral 1}}}\ radio surveys are close to the Milky Way disk's central H\;{\small{\MakeUppercase{\romannumeral 1}}}\ cavity\cite{Lockman_2016} ($|b|<10$\textdegree) and have kinematics that can be successfully modeled with a biconical outflow. As such, they are thought to originate in the disk of the Milky Way.  However, UV-detected FB HVCs seen at Galactic latitudes up to 55\textdegree\ (a projected distance of $\sim$10 kpc from the disk)  are not so clearly connected to the Milky Way disk and have never had their origin confirmed. Simulations have shown that it is difficult to accelerate a cool cloud in a galactic outflow without destroying it\cite{McCourt_2015, Scannapieco_2015, Schneider_2017, Zhang_2017}, as the cloud has to survive the shock and shear of the outflow. Instead, some of the high-latitude UV-detected FB HVCs may be halo clouds that pre-existed the formation of the bubbles.

\begin{figure}[!ht]
    \centering\textbf{Map of Metallicity Measurements}\par\medskip  %Title
     \includegraphics[width=\textwidth]{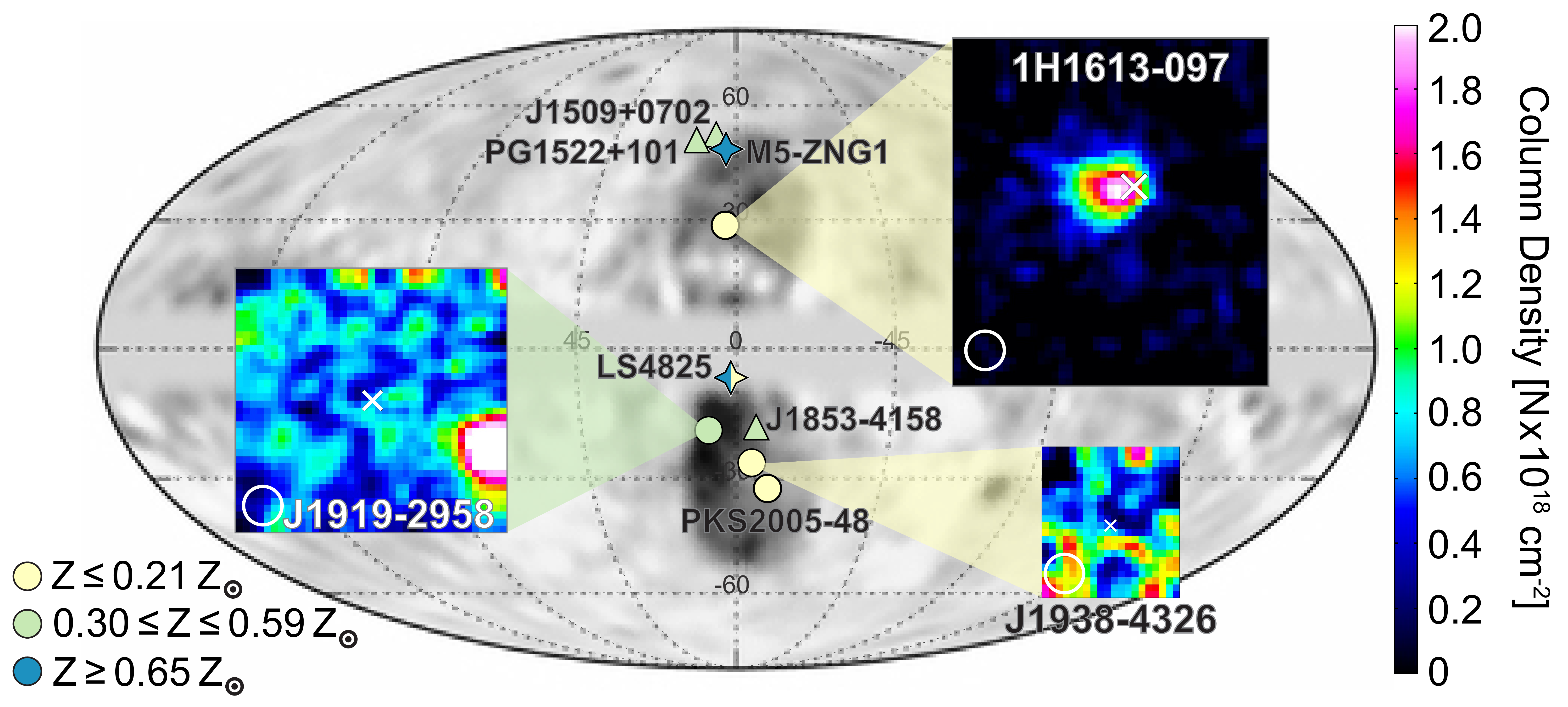}
\caption{Map of the Fermi Bubbles with UV sight lines from this sample plotted on top of an adapted 3--10 GeV residual intensity Fermi gamma-ray map\cite{Ackermann_2014} (grayscale is $-5$ to $10$ x\;$10^{7}$ GeV cm$^{-2}$ s$^{-1}$ sr$^{-1}$). Yellow, green, and blue symbols correspond to clouds with metallicities of $\leq21$\%, $30$\%$\leq Z\leq 59$\%, and $\geq65$\%\ solar, respectively. Circles represent background quasars, stars represent background stellar sight lines, and triangles represent lower-limit metallicity measurements from background quasars. We include the GBT H\;{\small{\MakeUppercase{\romannumeral 1}}}\ maps (when available) plotted to approximately the same angular scale for the sight lines. The circles in the bottom left of each subpanel indicate the 21 cm beam size and the white ``X'' indicates the location of the UV sight line. The colorscale used for the H\;{\small{\MakeUppercase{\romannumeral 1}}}\ maps represents the H\;{\small{\MakeUppercase{\romannumeral 1}}}\ column densities. The maps were made by integrating over velocities of $-201$ to $-150$ km~s$^{-1}$\ for 1H1613-097,  $-123$ to $-56$ km~s$^{-1}$\ for J1938-4326, and $65$ to $131$ km~s$^{-1}$\ for J1919-2958. The column colorbar units are in $N\times10^{18}$ cm$^{-2}$ where $N=1$, $3$, and $5$ for 1H1613-097, J1938-4326, and J1919-2958, respectively. }\label{figure:hi_maps}

\end{figure}

Metallicity measurements (heavy metal abundances) from H\;{\small{\MakeUppercase{\romannumeral 1}}}\ and UV data can directly constrain the origin of the FB HVCs, because clouds driven into the bubbles from the Galactic disk are expected to show solar (or super-solar) metallicities, whereas halo clouds should have sub-solar metallicities\cite{Wakker_1999, Richter_2001, Fox_2004}. We have conducted the first metallicity survey of FB HVCs by combining measurements of three newly-detected FB HVCs with measurements of nine clouds categorized as FB HVCs in the literature\cite{Keeney_2006, Bordoloi_2017, Savage_2017, Ashley_2020}. We calculate new ionization corrections and metallicities for the literature sample in all but one case (see the Methods section for more details). The three newly-detected FB HVCs each have UV spectra from the Cosmic Origins Spectrograph on the Hubble Space Telescope (HST/COS) as well as H\;{\small{\MakeUppercase{\romannumeral 1}}}\ maps and deep single-pointing H\;{\small{\MakeUppercase{\romannumeral 1}}}\ spectra from the Green Bank Telescope (GBT). In Figure~\ref{figure:hi_maps} we show the location of each sight line on a gamma-ray map of the Fermi Bubbles, including H\;{\small{\MakeUppercase{\romannumeral 1}}} maps of the FB HVCs with detected H\;{\small{\MakeUppercase{\romannumeral 1}}}\ emission when available. The H\;{\small{\MakeUppercase{\romannumeral 1}}}\ maps show the environment, morphology, and spatial size of each cloud. The FB HVC toward 1H1613-097 is a centrally concentrated, compact structure ($<$0.4\textdegree\ in size) while the new FB HVCs, J1919-2958 and J1938-4326 are tenuous and diffuse in H\;{\small{\MakeUppercase{\romannumeral 1}}}. These H\;{\small{\MakeUppercase{\romannumeral 1}}}\ maps are used to quantify the effects of beam-smearing on each H\;{\small{\MakeUppercase{\romannumeral 1}}}\ measurement (see the Methods section). We provide a list of each cloud in Table~\ref{table:main_metallicities} and detailed information for each cloud and sight line in the Methods Section.

\begin{figure}[!ht]
   \begin{center}
      \textbf{UV Absorption and HI Emission Profiles Used for Metallicity Measurements}\par\medskip  %Title
     \includegraphics[ width=0.80\textwidth]{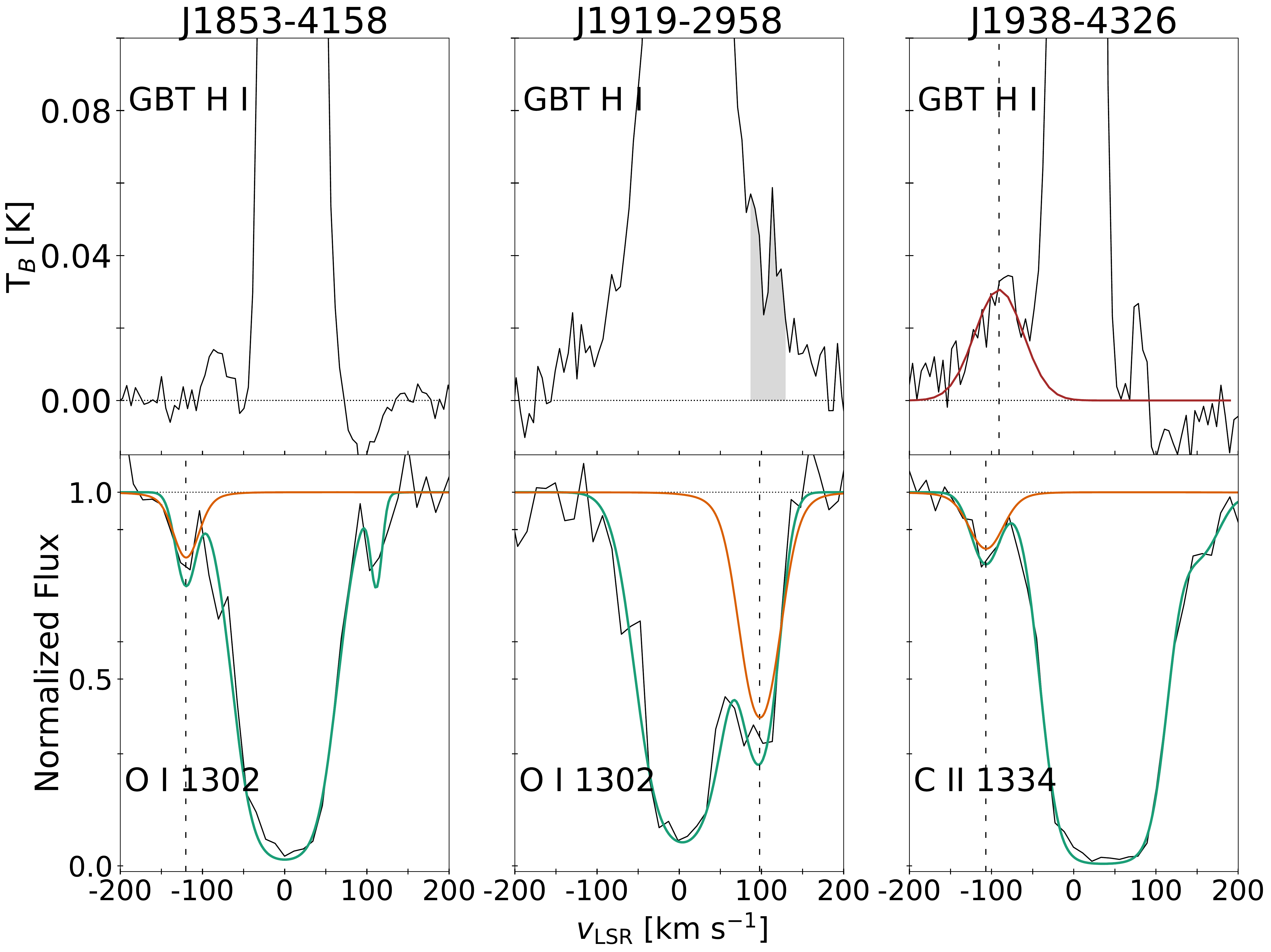}
    \end{center}
\caption{ H\;{\small{\MakeUppercase{\romannumeral 1}}}\ emission-line and UV absorption-line profiles used to calculate metallicities for each new sight line. Voigt fits to the UV components of each FB HVC are plotted on top of the data in orange. The full fit to all UV absorption components is plotted in green\cite{Ashley_2020}. No H\;{\small{\MakeUppercase{\romannumeral 1}}}\ emission associated with J1853-4158's FB HVC was detected at the sensitivity of the data. J1919-2958's H\;{\small{\MakeUppercase{\romannumeral 1}}}\ column limits were computed using the velocity range indicated by the grey shaded region after flipping the spectrum around its peak emission and subtracting the resulting spectrum from the original, as discussed in detail in the Methods section.  J1938-4326's Gaussian fit to the FB HVC H\;{\small{\MakeUppercase{\romannumeral 1}}}\ emission is indicated by a red line.}\label{figure:gauss_fit}
\end{figure}

Of all the absorption lines covered in the COS dataset (observed with the G130M and G160M gratings), O\;{\small{\MakeUppercase{\romannumeral 1}}}\ $\lambda$1302 and S\;{\small{\MakeUppercase{\romannumeral 2}}}\ $\lambda$1250, $\lambda$1253, $\lambda$1259 are the most useful for interstellar metallicity measurements because their ionization corrections are small at large H\;{\small{\MakeUppercase{\romannumeral 1}}}\ column densities (log\,$N_{\mathrm{HI}}\gtrsim18.5$ and $19.5$ for O\;{\small{\MakeUppercase{\romannumeral 1}}}\ and S\;{\small{\MakeUppercase{\romannumeral 2}}}, respectively; $N_{\mathrm{HI}}$ in cm$^{-2}$)\cite{Jenkins_2009,Bordoloi_2017}. Additionally, oxygen does not have large levels of dust depletion and even though sulfur has been shown to sporadically deplete (up to 1 dex in the dustiest clouds)\cite{Jenkins_2009}, sulfur is not typically thought to strongly deplete onto dust\cite{Savage_1996}. We therefore use these two ions for metallicity measurements. We have selected our sample to contain all known FB HVCs with detected O\;{\small{\MakeUppercase{\romannumeral 1}}}\ or S\;{\small{\MakeUppercase{\romannumeral 2}}}\ absorption and associated H\;{\small{\MakeUppercase{\romannumeral 1}}}\ detections. Another measurement included in our sample is a FB HVC toward J1938-4326, which has a C\;{\small{\MakeUppercase{\romannumeral 2}}}\ $\lambda$1334 detection and an associated H\;{\small{\MakeUppercase{\romannumeral 1}}}\ detection. J1938-4326's O\;{\small{\MakeUppercase{\romannumeral 1}}}\ absorption is blended with the background quasar's intrinsic Si\;{\small{\MakeUppercase{\romannumeral 3}}}\ absorption and it has no detected S\;{\small{\MakeUppercase{\romannumeral 2}}}\ absorption. Instead, we use C\;{\small{\MakeUppercase{\romannumeral 2}}}\ $\lambda$1334 for the metallicity tracer in this cloud as carbon is only weakly depleted \cite{Jenkins_2009}, although the ionization correction is substantial. 
We also incorporate PKS 2005-489's metallicity measurement from the literature into the sample\cite{Keeney_2006}. Neither O\;{\small{\MakeUppercase{\romannumeral 1}}}\ nor S\;{\small{\MakeUppercase{\romannumeral 2}}}\ detections exist for PKS 2005-489. Instead, its metallicity measurement was made by comparing its H\;{\small{\MakeUppercase{\romannumeral 1}}}\ column and detected ion abundances to those of Milky Way halo gas clouds with measured metallicities; therefore, all ions detected in this sight line were used to determine the FB HVC's metallicity\cite{Keeney_2006}. Additionally, we include FB HVCs with distinct O\;{\small{\MakeUppercase{\romannumeral 1}}}\ detections and H\;{\small{\MakeUppercase{\romannumeral 1}}}\ column density upper limits in our sample, providing lower limit metallicity measurements. We only use O\;{\small{\MakeUppercase{\romannumeral 1}}}\ $\lambda$1302 detections with H\;{\small{\MakeUppercase{\romannumeral 1}}}\ upper limits because (unlike other elements) oxygen has small ionization corrections for log\,$N_{\mathrm{HI}}\gtrsim18.5$ $^\mathrm{[}$\cite{Bordoloi_2017}$^\mathrm{]}$. In Figure~\ref{figure:gauss_fit} we plot the O\;{\small{\MakeUppercase{\romannumeral 1}}}\ or C\;{\small{\MakeUppercase{\romannumeral 2}}}\ absorption profiles and H\;{\small{\MakeUppercase{\romannumeral 1}}}\ single-pointing emission profile for the three newly-detected FB HVCs in our sample.

We find that four FB HVCs in our sample have low metallicities of $\le21$\%\ solar (one of which is a lower limit), five have metallicities between $\ge30$\% and $\ge59$\% solar, and three have near solar or supersolar metallicities of $\geq85$\% solar (see Table~\ref{table:main_metallicities}). The low metallicities ($\sim$ $20$\% solar) are indicative of Galactic halo clouds\cite{Wakker_1999, Richter_2001} and are similar to that measured for the X-ray component of the Fermi Bubbles using Suzaku X-ray data\cite{Kataoka_2013}. By comparison, disk clouds that come from the Galactic center should have solar or supersolar metallicities\cite{Afflerbach_1997, Rolleston_2000}.

Only one of our sight lines, LS 4825, probes the low-latitude portion of the Fermi Bubbles, close to the disk and central H\;{\small{\MakeUppercase{\romannumeral 1}}}\ cavity ($b=-6.63$\textdegree) where FB HVCs are expected to have near solar metallicities. LS~4825 has four FB HVCs which have distances between 7 and 21 kpc as evidenced by their absence in absorption spectra toward a closely aligned foreground star\cite{Savage_2017}. We find that the FB HVCs toward LS~4825 (LS~4825-1, -2, -3, -4 at $v_{0\:\mathrm{UV}}=92$, $-78$, $-205$, and  $-155$ km~s$^{-1}$, respectively) have metallicities of $\ge21$\%, 85$\pm^{41}_{37}$\%\, $\ge44$\%, and $\ge260$ \%\ solar. Two of these metallicities are near solar (LS 4825-2 and -4). Of the other two metallicity measurements, LS 4825-1 has an uncertain H\;{\small{\MakeUppercase{\romannumeral 1}}}\ measurement due to a complex H\;{\small{\MakeUppercase{\romannumeral 1}}}\ emission spectrum and potential narrow line O\;{\small{\MakeUppercase{\romannumeral 1}}}\ absorption (see the Methods section for details), and LS 4825-3 has a measurement based on an H\;{\small{\MakeUppercase{\romannumeral 1}}}\ upper limit. Therefore, it is possible that these two FB HVCs have a higher metallicities than reported.

As explained in detail in the Methods section, the metallicity measurements of the new FB HVCs are robust against beam-smearing effects that result from combining pencil-beam UV data with finite-beam radio data. Furthermore, we derive the time-dependent ionization corrections for the FB HVCs and find that a Galactic center Seyfert flare event 3.5 Myr ago does not have a significant effect on the clouds' ionization corrections\cite{Bland_Hawthorn_2013, Bland_Hawthorn_2019}. Neither of these effects change the result that the FB HVCs have metallicities ranging from  $<20$\% to $320$\%\ solar. These results challenge the previously accepted picture that all FB HVCs were driven out from the Galactic center with solar or even super-solar metallicity.

One potential explanation for the sub-solar metallicities observed in many of the FB HVCs is that the clouds originate in the Milky Way disk with solar metallicity and then become diluted to subsolar metallicities via metal mixing as they travel through the Fermi Bubbles to higher latitudes. However, as we discuss below, we find that metal mixing cannot explain FB HVC metallicities of $\mathrm{Z}\lesssim20\%$ solar. Mixing can occur when cold gas clouds traveling through a hot ambient medium leave a wake of stripped gas behind them. The stripped gas then mixes with the ambient gas reducing its temperature enough to allow for condensation of the hot medium onto the stripped cold cloudlets, resulting in a cloud with a metallicity between the cold cloudlets and hot plasma\cite{Gritton_2014, Gronke_2018}. In this scenario, FB HVCs originating from Milky Way's disk with solar (or super-solar) metallicity would need to be mixed down to subsolar metallicities ($\le20$\% in some cases) by low-metallicity hot plasma. The metallicity of the Fermi Bubbles' plasma is difficult to determine. The Fermi Bubble X-ray component has a measured metallicity of $\sim$20\%\ solar\cite{Kataoka_2013}; however, these measurements are based on limited photon statistics and are not necessarily indicative of the metallicity in gamma-ray component of the bubbles. The gamma-ray-emitting plasma is likely enriched with metals from the Milky Way's ionized interstellar medium and from the shocked hot Milky Way halo component as the bubbles expand into the halo. The interstellar medium has near-solar metallicities and the hot halo has a relatively high metallicity of $\sim$60\%\ solar\cite{Gritton_2014, Miller_2016a, Henley_2017} so neither source is expected to reduce the Fermi Bubble plasma's metallicity to $\le20$\%\ solar. Additionally, current models of cold clouds moving through the Milky Way halo find that it takes tens of Myr for significant portions of the ambient hot halo to condense onto the cloudlets\cite{Armillotta_2016,Marasco_2011}; this timescale would be even longer in the Fermi Bubbles, which are approximately twice the temperature of the Milky Way halo and therefore take longer to cool and condense\cite{Miller_2016b}. As such, given that the Fermi Bubbles are only 3-6 Myr old$^{\mathrm[}$\cite{Guo_2012, Mou_2014, Miller_2016b, Fox_2020}$^{\mathrm]}$, cold cloudlets would not have sufficient time to condense a significant amount of the plasma, leaving them with a relatively unchanged metallicity. Considering all of these effects, we conclude it is unlikely that mixing can explain FB HVCs with Z$\lesssim$20\%\ solar. 

We suggest an alternative explanation. FB HVCs have a large range in metallicities because they have two origins: the Milky Way's disk and its halo. In this picture, one population of FB HVCs is composed of the low-latitude FB HVCs that originate in the disk of the Milky Way and are accelerated away from the Galactic center by the same mechanism that created the Fermi Bubbles. These cool gas clouds could be accelerated by the hot gas flow and eventually become eroded as their mass transfers from a cool to hot phase\cite{Zhang_2017, Schneider_2020}. The remnants of many of these clouds likely become too dispersed, ionized, and hot to be seen in the UV as they move to higher latitudes. The second population of FB HVCs is composed of pre-existing halo clouds that are shocked and accelerated as the Fermi Bubbles expand into the halo. As such, they represent the medium into which the Fermi Bubbles expand, rather than material carried out by the Milky Way's nuclear outflow. 

The division between the disk and halo clouds does not occur at a single latitude, as evidenced by the FB HVC toward M5-ZNG1 (high metallicity at a high latitude of 47\textdegree). This outlier suggests that the outflow is complex and that some disk clouds may be able to survive to higher latitudes in the bubbles. While it is not clear how these clouds survive the shock and shear of the Fermi Bubbles, a range of ideas have been suggested from internal magnetic fields suppressing the destruction of the clouds to radiative cooling increasing a cloud's lifetime in a hot outflow\cite{McCourt_2015, Scannapieco_2015, Schneider_2017, Zhang_2017}. In any case, the UV-detected FB HVCs reported in this paper have a large range of metallicities from $<20$\% solar to $320$\%\ solar, a result that requires a revision to the canonical picture of FB HVCs and indicates that biconical outflow models need to be refined to account for the clouds' dual origins.

Our new results on the metallicity and origin of FB HVCs reveal the properties of a nuclear outflow in more detail than is possible in any other spiral galaxy. As such, the Fermi Bubbles serve as an important analog for extragalactic nuclear outflows, which can typically only be studied with one-dimensional down-the-barrel pointings.
In general, galactic outflows and outward-moving shocks are thought to prevent inflowing cool gas clouds from reaching the disk of a galaxy, thus quenching the star-formation process\cite{Birnboim_2003}. While the Fermi Bubbles are not likely strong enough to quench the Milky Way\cite{Cattaneo_2007, Beckmann_2017}, they remain important for understanding how nuclear outflows entrain and sweep up cool halo gas, thus circulating matter between the disk and halos of galaxies.

\clearpage
\section*{Figures}

\clearpage
\section*{Tables}
\begin{table}[hb!]
\begin{center}
\caption{\textbf{Sample and Metallicity Measurements}}\label{table:main_metallicities}
\resizebox{0.8\columnwidth}{!}{%
\begin{tabular}{lcccccccc}
\hline \hline
Reference & Sight line & $l$ & $b$ &  Spectral & X$^{i}$ & $v_{0\:\mathrm{UV}}$ & Z \\
 & & (\textdegree) & (\textdegree) &  Type &  & (km~s$^{-1}$) & (Z$_{\odot}$)\\\hline
\noalign{\vskip 0.5ex}
\multicolumn{8}{c}{UV and H\;{\small{\MakeUppercase{\romannumeral 1}}}\ detections } \\\hdashline
TW & J1919-2958 & $8.18$  &  $-18.77$   &  Quasar &  O\;{\small{\MakeUppercase{\romannumeral 1}}} & $97.8\pm7.7$ & $[0.30,0.54]$ \\
TW & J1938-4326 & $355.47$  &  $-26.41$  & Quasar &  C\;{\small{\MakeUppercase{\romannumeral 2}}} & $-107.1\pm8.2$ & $[0.010, 0.10]$\\
\citen{Bordoloi_2017} & 1H1613-097 & $3.53$ & $28.46$ & Quasar & O\;{\small{\MakeUppercase{\romannumeral 1}}} & $-163\pm1$ & $0.19\pm0.10$\\
\citen{Savage_2017} & LS 4825-1 & $1.67$ & $-6.63$ & B1 Ib-II & O\;{\small{\MakeUppercase{\romannumeral 1}}} & $92.1\pm0.8$ & $\ge0.21$ \\
\citen{Savage_2017} & LS 4825-2 & $1.67$ & $-6.63$ & B1 Ib-II &  S\;{\small{\MakeUppercase{\romannumeral 2}}} & $-78\pm1$ & $0.85\pm^{0.41}_{0.37}$ \\
\citen{Zech_2008} & M5-ZNG1 & $3.86$ & $46.79$ & sdO & O\;{\small{\MakeUppercase{\romannumeral 1}}} & $-124\pm7$ & $3.2\pm0.8$\\
\citen{Keeney_2006} & PKS 2005-489 & $350.37$ & $-32.60$ & Quasar &  ... & $168\pm10$ & $\le0.20$\\\hdashline\noalign{\vskip 0.5ex}
\multicolumn{8}{c}{ UV detections and H\;{\small{\MakeUppercase{\romannumeral 1}}}\ limits} \\\hdashline
TW & J1853-4158 & $354.36$  &  $-18.04$ &  Quasar & O\;{\small{\MakeUppercase{\romannumeral 1}}} & $-120.2\pm8.4$ & $\ge0.59$ \\
\citen{Bordoloi_2017} & J1509-0702 & $7.80$  &  $51.61$ &  Quasar & O\;{\small{\MakeUppercase{\romannumeral 1}}} & $-115.2\pm7.8$ & $\ge0.59$ \\
\citen{Savage_2017} & LS 4825-3 & $1.67$ & $-6.63$ & B1 Ib-II & O\;{\small{\MakeUppercase{\romannumeral 1}}} & $-205\pm2$ & $\ge0.44$ \\
\citen{Savage_2017} & LS 4825-4 & $1.67$ & $-6.63$ & B1 Ib-II &  O\;{\small{\MakeUppercase{\romannumeral 1}}} & $-155\pm1$ & $\ge2.6$ \\
\citen{Bordoloi_2017} & PG 1522+101 & $14.89$  &  $50.12$ &  Quasar & O\;{\small{\MakeUppercase{\romannumeral 1}}} & $-99.8\pm8.3$ & $\ge0.30$ \\

\hline
\end{tabular}%
}
\end{center}
\vspace{-20pt}
\caption*{The references give the source for the UV and H\;{\small{\MakeUppercase{\romannumeral 1}}}\ measurements (TW=This Work). The Galactic latitude and longitude are given as $l$ and $b$, respectively. The type of each background source is listed as quasar or the stellar spectral type.  X$^{i}$ represents the ion used for metallicity calculations. The UV Voigt fit local standard of rest (LSR) velocity centroid is $v_{0\:\mathrm{UV}}$. $Z$ represents the corrected elemental abundances in linear form (for further details on $Z$ calculations, see the Methods Section and Extended Data Fig.~\ref{table:metallicities}). J1919-2958's metallicity is represented by a range due to stray radiation effects (see the Methods section for details). LS 4825 has four detected clouds, listed separately. The M5-ZNG1's measurement applies to two absorption components that are blended in H\;{\small{\MakeUppercase{\romannumeral 1}}}. We treat them as a single cloud because only an average metallicity measurement is possible. PKS 2005-489's metallicity measurement is a literature measurement made by comparing UV ion and Lyman measurements to other Milky Way HVCs\cite{Keeney_2006}. J1938-4326's metallicity limit includes an estimate for dust depletion and is a range due to multiple H\;{\small{\MakeUppercase{\romannumeral 1}}}\ measurements, as explained in detail in Supplementary Information Section 1.}
\end{table}

\clearpage

%TC:endignore
\section*{Methods}

\subsection*{Sample} 
Our full sample consists of nine sight lines passing through or near the Fermi Bubbles. Three of the nine sight lines are new quasar sight lines\cite{Ashley_2020} and the other six are taken from the literature\cite{Keeney_2006, Zech_2008, Bordoloi_2017, Savage_2017}. We have UV O\;{\small{\MakeUppercase{\romannumeral 1}}}\ or S\;{\small{\MakeUppercase{\romannumeral 2}}}\ absorption and H\;{\small{\MakeUppercase{\romannumeral 1}}}\ emission measurements for five FB HVCs. We also have O\;{\small{\MakeUppercase{\romannumeral 1}}}\ absorption measurements and measure the H\;{\small{\MakeUppercase{\romannumeral 1}}}\ limits for five more FB HVCs. Additionally, we have one FB HVC with C\;{\small{\MakeUppercase{\romannumeral 2}}}\ absorption and H\;{\small{\MakeUppercase{\romannumeral 1}}}\ emission measurements and another FB HVC with a UV-based literature metallicity measurement.  The UV data for all nine sight lines has been previously calibrated and analyzed\cite{Keeney_2006,Zech_2008, Bordoloi_2017, Savage_2017, Ashley_2020}. We make new measurements for the UV absorption or H\;{\small{\MakeUppercase{\romannumeral 1}}}\ emission associated with three literature FB HVCs (J1509-0702, PG 1522+101, and LS 4825; see the ``UV Ion and H\;{\small{\MakeUppercase{\romannumeral 1}}}\ Column Densities of the Literature Sample'' section below). Most of the FB HVC's in our sample have not had their metallicities previously calculated and many of the metallicity calculations in the literature do not include an ionization correction, an important correction for these clouds that are exposed to the Galactic, extragalactic, and Seyfert flare ionizing radiation fields, in addition to cosmic rays. Table~\ref{table:main_metallicities} lists information for each sight line and associated cloud(s). The names of several of the targets are abbreviated throughout the paper as follows: J1853-4158=UVQS J185302.65-415839.6, J1919-2958=UVQS J191928.05-295808.0, J1938-4326=UVQS J193819.59-432646.3, and J1509-0702= SDSS J150928.29+070235.8.  

\subsection*{Observations and Data Reduction for Three New  Sight Lines}\label{section:observations}
The three new targets were selected from the UV-bright QSO Survey (UVQS) catalog\cite{Monroe_2016} and were observed under HST Program ID 15339 (PI: A. Fox) and GBT Programs  GBT18A-221 (PI: A. Fox) and GBT20B-444 (PI: T. Ashley). For each new sight line we obtained both GBT single-pointing spectra and GBT H\;{\small{\MakeUppercase{\romannumeral 1}}}\ maps. The single-pointing spectra were used for metallicity measurements and the maps were used for beam-smearing calculations and to ascertain the FB HVCs' gaseous environments.

The Voigt fit parameters of the new FB HVCs' UV ions used for metallicity measurements are listed in Extended Data Fig.~\ref{table:UV_HI_fits}. We note that the FB HVC toward J1919-2958 may be composed of two strongly blended absorption features separated by $\lesssim$20 km~s$^{-1}$, the spectral resolution of COS. Without clearly separated components in any other UV-detected ions\cite{Ashley_2020} we are unable to confirm that there are two absorption features contributing to the FB HVC. Therefore, we treat the absorption and emission associated with the FB HVC as one cloud.

The GBT single-pointing H\;{\small{\MakeUppercase{\romannumeral 1}}}\ spectra were calibrated using standard techniques\cite{Boothroyd_2011}. This includes a correction for stray radiation, defined as any 21-cm H\;{\small{\MakeUppercase{\romannumeral 1}}}\ emission that enters the receiver through the telescope’s sidelobes, rather than via the main beam.  This complicates our derivation of an accurate $N_{\mathrm{HI}}$ for J1919-2958, where a GBT sidelobe lies near the Galactic plane, picking up stray emission at $v_{\rm{LSR}}<125$ km~s$^{-1}$\ for most of the time when this quasar is observable from Green Bank. For J1919-2958, stray radiation dominates the detected signal over the velocity range of the UV absorption and the standard correction technique yields highly uncertain, and in some cases unphysical results.  Because stray radiation must always have a positive amplitude, for J1919-2958 we use the scans with minimum H\;{\small{\MakeUppercase{\romannumeral 1}}}\ emission in the uncorrected data to derive a strict upper limit and a range on the possible values of the H\;{\small{\MakeUppercase{\romannumeral 1}}}\ column density.

Toward J1938-4326 the entire observed HI spectrum was decomposed into Gaussian components using the minimum number of components needed to obtain a fit with low residuals. The resulting Gaussian fit parameters are listed in Extended Data Fig.~\ref{table:UV_HI_fits}.

We were unable to fit accurate Gaussian components to J1919-2958's H\;{\small{\MakeUppercase{\romannumeral 1}}}\ spectrum due to strong blending with the Milky Way's H\;{\small{\MakeUppercase{\romannumeral 1}}}\ emission. To measure  J1919-2958's H\;{\small{\MakeUppercase{\romannumeral 1}}}\ column densities, we assume that the Milky Way disk's H\;{\small{\MakeUppercase{\romannumeral 1}}}\ emission has symmetric positive and negative velocity wings and there is only one FB HVC along the line of sight.  The H\;{\small{\MakeUppercase{\romannumeral 1}}}\ spectrum is inverted along its velocity axis about the velocity of the peak intensity (close to zero) and subtracted from the original spectrum. The remaining ``flipped-and-subtracted" spectrum is shown in Extended Data Fig.~\ref{figure:1919_hi_flip}. We then integrate the remaining emission with a velocity near that of the UV-detected FB HVC velocity range ($v_{0\:\mathrm{UV}}=97.8\pm7.7$ km~s$^{-1}$, $b=25.9\pm2.5$ km~s$^{-1}$) and calculate log~$N_{\mathrm{HI}}$ using Equation~\ref{equation:nhi}. For this method, we use two velocity ranges for J1919-2958's H\;{\small{\MakeUppercase{\romannumeral 1}}}\ integration: $82.7-133.4$ km~s$^{-1}$\ and $107.0-133.4$ km~s$^{-1}$.  The velocity ranges were chosen to include both or one of the apparent H\;{\small{\MakeUppercase{\romannumeral 1}}}\ Gaussian-like features in the ``flipped-and-subtracted" spectrum near the velocity range of the O\;{\small{\MakeUppercase{\romannumeral 1}}}\ absorption feature (see Extended Data Fig.~\ref{figure:1919_hi_flip}). The larger range is inclusive of all potential emission associated with the FB HVC and the stray radiation, making the column density an upper limit, while the smaller range excludes any emission that likely contains strong stray radiation, providing a lower limit for the H\;{\small{\MakeUppercase{\romannumeral 1}}}\ column density. We calculate the error on the area as the average root-mean-square (rms) of the noise (0.024 K; measured from emission-free channels) times the channel width and the square root of the number of channels.

The H\;{\small{\MakeUppercase{\romannumeral 1}}}\ column density for J1919-2958 and 1938-4326 was calculated using the standard relation: 
\begin{equation}\label{equation:nhi}
N_{\mathrm{HI}}[\mathrm{cm}^{-2}]=1.823*10^{18} \int T_B(v)~ dv\:[\mathrm{K}\:\mathrm{km}~\mathrm{s}^{-1}]
\end{equation}
where $T_B(v)$ is the velocity-dependent brightness temperature. For the Gaussian fit to J1938-4326's FB HVC, the integral reduces to $\int T_B(v)~ dv\:[\mathrm{K}\:\mathrm{km}~\mathrm{s}^{-1}] = 1.064\: h\:\mathrm{FWHM}\:[\mathrm{K}\:\mathrm{km}~\mathrm{s}^{-1}]$, where $h$ is the height of the Gaussian and FWHM is the full width of the Gaussian at half maximum. A check on J1938-4326's H\;{\small{\MakeUppercase{\romannumeral 1}}}\ column density measurement is conducted in Supplementary Information Section 2 using the flip-and-subtract method described above. For J1853-4158, we calculate the H\;{\small{\MakeUppercase{\romannumeral 1}}}\ column upper limit as that which is given by three times the rms over the detected O\;{\small{\MakeUppercase{\romannumeral 1}}}'s FWHM:\\
\begin{equation}\label{Equation:HI_limit}
N_{\mathrm{HI}}~\le~3{*}1.823{*}10^{18}\:rms\:\Delta v\:\sqrt{n}, 
\end{equation}
where $\Delta v$ is the channel width in km~s$^{-1}$, and $n$ is the number of H\;{\small{\MakeUppercase{\romannumeral 1}}}\ spectrum channels in the velocity range of the detected O\;{\small{\MakeUppercase{\romannumeral 1}}}'s FWHM.
The H\;{\small{\MakeUppercase{\romannumeral 1}}}\ column densities for the three new FB HVCs are listed in Extended Data Fig.~\ref{table:UV_HI_fits}.

GBT H\;{\small{\MakeUppercase{\romannumeral 1}}}\ 21 cm emission maps around the central quasars were obtained in addition to single-pointing spectra for the three new sight lines. These frequency-switched data were calibrated using standard techniques, smoothed to $\sim3$ km~s$^{-1}$, and corrected for stray radiation\cite{Boothroyd_2011}. All data from the XX polarization in the GBT maps are discarded due to spurious spectral features, as discussed in the Supplementary Information Section~3 (see also Supplementary Figure~1).

We present the GBT H\;{\small{\MakeUppercase{\romannumeral 1}}}\ moment zero map for J1919-2958 and J1938-4326 in Figure~\ref{figure:hi_maps} (approximately 1.0\textdegree$\times$1.0\textdegree\ for J1919-2958 and 0.5\textdegree$\times$0.5\textdegree\ for J1938-4326). We do not include J1853-4158's H\;{\small{\MakeUppercase{\romannumeral 1}}}\ map because its single-pointing spectrum does not show an H\;{\small{\MakeUppercase{\romannumeral 1}}}\ detection. We also include in Figure~\ref{figure:hi_maps} a deep GBT H\;{\small{\MakeUppercase{\romannumeral 1}}}\ map of the 1H1613-097 field that was obtained under GBT programs GBT20A-253 and GBT17B-015 ($\sim$1.2\textdegree\ in width). These maps were created by visually inspecting the spectrum extracted from the data cube integrated over one beam around the sight line (using the H\;{\small{\MakeUppercase{\romannumeral 1}}}\ single-pointing detections as a guide). We then integrate channels with potential H\;{\small{\MakeUppercase{\romannumeral 1}}}\ emission from the cloud of interest (velocities of $65$ to $131$ km~s$^{-1}$\ for J1919-2958, $-123$ to $-56$ km~s$^{-1}$\ for J1938-4326, and $-201$ to $-150$ km~s$^{-1}$\ for 1H1613-097). 

\subsection*{Beam-Smearing Analysis}

We use the new FB HVCs' GBT H\;{\small{\MakeUppercase{\romannumeral 1}}}\ maps to investigate small-scale structure by quantifying the level by which beam-smearing affects our measurements. Since we also have 1H1613-097's GBT map, we included it in our beam-smearing analysis. To quantify beam-smearing effects, we extract spectra from the GBT maps within four equally-spaced beam-sized pointings, each centered a half beamwidth away from the central quasar sight line, and one pointing centered on the sight line itself. We perform a boxcar smoothing of each spectrum by three channels and fit Gaussians to each FB HVC. We then calculate each cloud's log $N_{\mathrm{HI}}$ in each pointing and take the largest difference between log $N_{\mathrm{HI}}$ from the central pointing and the four surrounding pointings as a beam smearing error, $\epsilon_{\mathrm{Beam}}$. We compile the beam smearing errors and standard deviation of both the H\;{\small{\MakeUppercase{\romannumeral 1}}}\ velocity centroids and FWHM of fits in Supplementary Table~1. The beam-smearing errors are added to the H\;{\small{\MakeUppercase{\romannumeral 1}}}\ column-density errors in quadrature when calculating the metallicity of each cloud. We note that due to the stray radiation problems discussed above in the Observations and Data Reduction section, J1919-2958's beam smearing error could be affected by stray radiation. However, the resulting  $\epsilon_{\mathrm{Beam}}$ of 0.10 dex is a comparable to other beam-smearing error estimates on these angular scales in the literature\cite{Fox_2018}.

The other literature sight lines either do not have H\;{\small{\MakeUppercase{\romannumeral 1}}}\ emission maps available (LS 4825, J1509-0702, and PG 1522+101) or have their H\;{\small{\MakeUppercase{\romannumeral 1}}}\ columns measured using Lyman lines from FUSE data with an infinitesimal beamsize (M5-ZNG1 and PKS 2005-489). For LS 4825's H\;{\small{\MakeUppercase{\romannumeral 1}}}\ measurements, we add in quadrature an assumed beam-smearing error of 0.15 dex to H\;{\small{\MakeUppercase{\romannumeral 1}}}\ columns in metallicity calculations, a reasonable assumption for GBT and HST UV derived metallicities\cite{Savage_2017, Bordoloi_2017}.

\subsection*{UV Ion and H\;{\small{\MakeUppercase{\romannumeral 1}}}\ Column Densities of the Literature Sample} 

For each cloud in the literature sample we use their respective UV ion and H\;{\small{\MakeUppercase{\romannumeral 1}}}\ column densities given in each reference, with the exception of LS 4825-1/2, J1509-0702, and PG 1522+101. We have listed the UV and H\;{\small{\MakeUppercase{\romannumeral 1}}}\ fit parameters used for metallicity measurements for the literature sample in Extended Data Fig.~\ref{table:UV_HI_fits_literature}.  \looseness=-2

LS 4825-1 has overlapping narrow and wide O\;{\small{\MakeUppercase{\romannumeral 1}}}\ components reported in the literature at 90.9 and 93.3 km~s$^{-1}$ $^{[}$\cite{Savage_2017}$^{]}$. In order to obtain a metallicity for LS 4825-1, we add the columns of both components and use the their average $v_{0\:\mathrm{UV}}$ and larger b-value of the two throughout the paper. We also use VPFIT to recalculate the O\;{\small{\MakeUppercase{\romannumeral 1}}}\ columns for two sources with apparent optical depth measurements in the literature from strongly blended absorption, J1509-0702 and PG~1522+101. For all other literature sources, we use the reported UV ion column densities. For M5-ZNG1 and 1H1613-097, we also use the literature reported H\;{\small{\MakeUppercase{\romannumeral 1}}}\ column densities.

For the FB HVC's with H\;{\small{\MakeUppercase{\romannumeral 1}}}\ limits, we calculate the H\;{\small{\MakeUppercase{\romannumeral 1}}}\ column upper limit from GBT (J1853-4158; LS 4825-3/4), LAB\cite{Kalberla_2005} (J1509-0702), or Green Bank 140' (PG~1522+101) spectra.  The H\;{\small{\MakeUppercase{\romannumeral 1}}}\ column upper limit is calculated two ways: using Equation~\ref{Equation:HI_limit} with the rms and using Equation~\ref{equation:nhi} with three times the H\;{\small{\MakeUppercase{\romannumeral 1}}}\ emission integrated over the velocity range of the respective O\;{\small{\MakeUppercase{\romannumeral 1}}}\ absorption's FWHM. From both these methods, the higher column density is used as a conservative H\;{\small{\MakeUppercase{\romannumeral 1}}}\ upper limit.

LS 4825's H\;{\small{\MakeUppercase{\romannumeral 1}}}\ spectra has strongly blended multi-component positive and negative velocity emission, making it difficult to fit a unique set of Gaussians to the spectrum. Additionally, the negative velocity H\;{\small{\MakeUppercase{\romannumeral 1}}}\ component between $-110$ and $-40$ km~s$^{-1}$\ is reported to have one strong emission peak centered at $-70.3$ km~s$^{-1}$\ and one weak emission peak centered at $-88.6$ km~s$^{-1}$. However, Mg\;{\small{\MakeUppercase{\romannumeral 1}}} and C\;{\small{\MakeUppercase{\romannumeral 1}}}, both of which trace cool gas, have absorption components of nearly equal magnitude at similar negative velocities (see Extended Data Fig.~\ref{figure:LS4825_HI})\cite{Savage_2017}. Since LS 4825's spectrum is so complex, we have decided to re-measure the H\;{\small{\MakeUppercase{\romannumeral 1}}}\ emission for the FB HVCs. 

For LS 4825's negative velocity FB HVC component, we estimate the low-velocity Milky Way H\;{\small{\MakeUppercase{\romannumeral 1}}}\ component's contribution to the negative high-velocity components by using HD 167402's H\;{\small{\MakeUppercase{\romannumeral 1}}}\ spectrum as a ``foreground" emission model. We first decompose HD 167402's full H\;{\small{\MakeUppercase{\romannumeral 1}}}\ spectrum into Gaussians with minimum residuals. Then, using these Gaussians as a model, we remove HD 167402's negative-velocity emission peaks that are not likely associated with the low-velocity Milky Way H\;{\small{\MakeUppercase{\romannumeral 1}}}\ component ($v_{0\mathrm{HI}}$ of $-51.5$ and $-73.9$ km~s$^{-1}$, FWHM of $18.8$ and $45.9$ km~s$^{-1}$, and peak amplitudes of $0.17$ and $0.10$ K, respectively; see Extended Data Fig.~\ref{figure:LS4825_HI}). HD 167402's residual  low-velocity Milky Way H\;{\small{\MakeUppercase{\romannumeral 1}}}\ emission has a slope that matches that of LS 4825's at negative velocities with only small variations, indicating that HD 167402's residual emission is a suitable model for LS 4825's low-velocity H\;{\small{\MakeUppercase{\romannumeral 1}}}\ emission at negative velocities. We subtract HD 167402's residual H\;{\small{\MakeUppercase{\romannumeral 1}}}\ spectrum from LS 4825's and fit LS 4825's remaining negative-velocity H\;{\small{\MakeUppercase{\romannumeral 1}}}\ components with Gaussians using the UV absorption components as a guide and the minimum number of components needed to obtain a low-residual fit. The fit was accomplished with two Gaussians of nearly equal strength, matching the Mg\;{\small{\MakeUppercase{\romannumeral 1}}} and C\;{\small{\MakeUppercase{\romannumeral 1}}} absorption trends (see Extended Data Fig.~\ref{figure:LS4825_HI}). We assume the $-85.8\pm1.7$ km~s$^{-1}$\ H\;{\small{\MakeUppercase{\romannumeral 1}}}\ component is associated with the $-78\pm1$ km~s$^{-1}$\ S\;{\small{\MakeUppercase{\romannumeral 2}}}\ absorption because of their adjacent velocities. While UV absorption and H\;{\small{\MakeUppercase{\romannumeral 1}}}\ emission lines from the same cloud should have approximately aligned velocity centroids, the size difference between the GBT 9.1$'$ beam and the \emph{HST}/STIS pencil beam could lead to different spatial coverage of a cloud with internal velocity structure, and therefore misaligned velocity centroids in the spectra.

The positive velocity Milky Way emission for HD 167402 and LS 4825 are significantly different; therefore, this method only works for negative velocity clouds. For the positive velocity FB HVC, we chose to integrate the H\;{\small{\MakeUppercase{\romannumeral 1}}}\ emission within the O\;{\small{\MakeUppercase{\romannumeral 1}}}\ absorption's FWHM of 19.8 km~s$^{-1}$\ and velocity centroid of $92.1\pm0.8$ km~s$^{-1}$.

\subsection*{Metallicity Measurements}\label{section:metallicity}

We calculate the metallicities from the ion abundances [X$^i$/H] as:

\begin{align}
\begin{aligned}\label{equation:XiH}
\left[\frac{{\rm X}}{\rm{H}}\right]=\left[\frac{{\rm X}^i}{\rm{H}}\right]+IC({\rm X}),\ \ \ \ \ \ \ \ \ \ \ \ \\
\mathrm{where}\ \  \left[\frac{{\rm X}^i}{\rm{H}}\right] = \mathrm{log}\left(\frac{N_{\mathrm{X}^i}}{N_\mathrm{H\;I}}\right) - \mathrm{log}\left(\frac{\rm{X}}{\rm{H}}\right)_\odot
\end{aligned}
\end{align}
where X is the element under study, X$^i$ is the observed ion of that element, IC is the ionization correction, $N_{\mathrm{X}^i}$ and $N_\mathrm{H\;I}$ are the ionic and atomic hydrogen column densities, respectively, and log\,(X/H)$_\odot$ is the solar abundance of that element\cite{Asplund_2009}. The ionization corrections are important for FB HVCs given their location close to the Galactic center and within the cone of a potential past Seyfert flare\cite{Bland_Hawthorn_2013, Bland_Hawthorn_2019} exposes them to intense radiation fields. Furthermore, their low $N$(H\;{\small{\MakeUppercase{\romannumeral 1}}}) values are indicative of high levels of ionization.

We derive the ionization corrections for the new FB HVCs by performing a suite of photoionization models using \texttt{CLOUDY} v.17.02 \cite{Ferland_2017}. The \texttt{CLOUDY} models are calculated for both an equilibrium case and a time-dependent case to explore the effects of a potential Seyfert flare\cite{Bland_Hawthorn_2013, Bland_Hawthorn_2019} on the ionization corrections.  Since J1919-2958 has two H\;{\small{\MakeUppercase{\romannumeral 1}}} column densities, we have run two sets of \texttt{CLOUDY} models for this cloud; these models are compared in the Supplementary Information Section 4. To increase our sample for the time-dependent models, we also include the FB HVC towards 1H1613-097\cite{Bordoloi_2017}. We discuss the ionization corrections for the rest of the sample below.

\subsection*{Equilibrium Models}\label{subsec:ioneq}

We first model the time-independent ICs for an ionization equilibrium state in \texttt{CLOUDY}. Physically these models describe the steady-state ionization conditions of clouds existing in the halo before a Seyfert flash event. These models are used as input parameters for the time-dependent models and are later used to confirm whether the time-dependent models strongly affect the present-day ionization corrections.

We adopt a plane-parallel geometry with a slab of uniform density illuminated on both sides by a position-dependent combined Galactic and extragalactic radiation field\cite{Bland_Hawthorn_1999, Fox_2005, Fox_2014}. The field has a normalization scaled to a distance of 10 kpc along the northern Fermi Bubble line of sight to 1H1613-097\cite{Bordoloi_2017}. This field is also suitable for J1919-2958 and J1938-4326 given that the radiation field models are comparatively symmetric up to 10 kpc\cite{Fox_2014}, as well as the similar absolute latitude of 1H1613-097 ($b=28.5^\circ$) with J1919-2958 ($b=-18.77^\circ$) and J1938-4326 ($b=-26.41^\circ$). Additionally, we include the cosmic ray background in our radiation field\cite{Indriolo_2007}. 

For each FB HVC, we construct a grid of \texttt{CLOUDY} models over a range of hydrogen number densities (log $n_\mathrm{H}$) between $-$2 and 1 in 0.5 dex intervals, with the metallicity fixed to the ion abundance values presented in Extended Data Fig.~\ref{table:metallicities}. The model for each cloud was run until their respective measured H\;{\small{\MakeUppercase{\romannumeral 1}}}\ column densities were reached. The value of log $n_\mathrm{H}$ was determined by comparing the observed column density ratio [Si\;{\small{\MakeUppercase{\romannumeral 3}}}/Si\;{\small{\MakeUppercase{\romannumeral 2}}}] to the model-inferred ratio at each step in the grid of hydrogen densities, and interpolating to find the best match (see Supplementary Fig.~2 for the observed Si\;{\small{\MakeUppercase{\romannumeral 2}}} and Si\;{\small{\MakeUppercase{\romannumeral 3}}} absorption profiles). Si\;{\small{\MakeUppercase{\romannumeral 2}}}\ absorption associated with J1938-4326's FB HVC was not detected. We instead use the Si\;{\small{\MakeUppercase{\romannumeral 3}}}/C\;{\small{\MakeUppercase{\romannumeral 2}}}\ ratio assuming that both ions have the same abundance relative to solar. 
Once a value for log $n_\mathrm{H}$ was determined for each FB HVC, the model was run again for that specific value in order to obtain exact model results.

\subsubsection*{Non-Equilibrium Time-dependent Models}\label{Section:time-dependent_models}

There is the strong evidence that a $\sim$3.5 Myr old Seyfert flare from the Galactic center impacted the Magellanic Stream and resulted in enhanced present-day H$\alpha$ emission and UV ionization levels in the stream\cite{Bland_Hawthorn_2013,Bland_Hawthorn_2019, Fox_2020}. Since the FB HVCs are approximately ten times closer to the Galactic center than the Magellanic Stream, there may also be enduring ionization effects from the Seyfert flare on the FB HVCs.  We use \texttt{CLOUDY} to follow the evolution of the FB HVCs near a time-variable incident radiation field. In this time-dependent non-equilibrium model, ionization and recombination are not in balance and thus a net rate of change for an ion's density occurs\cite{Chatzikos_2015, Ferland_2017}.

The shape of the flare was modeled using the AGN spectrum built into \texttt{CLOUDY}, which consists of a multi-component continuum characterized by a rising power law with a high energy exponential cutoff and a ``big blue bump" component peaking at $\approx$1 Ryd. The intensity of the flare is specified by the surface flux of hydrogen-ionizing photons, $\Phi$(H), striking the illuminated face of the cloud. The cloud distances from the Galactic center range from 2.7 to 4.1 kpc, which translates to $10^{9.25} < \Phi(\mathrm{H}) < 10^{9.75}$ photons cm$^{-2}$ s$^{-1}$ for an Eddington fraction $f_E$ = 0.3$^{[}$\cite{Bland_Hawthorn_2019}$^{]}$. We run our models for the range 8 $<$ log $\Phi(\mathrm{H})$ $<$ 10 to bracket the expected range of values\cite{Bland_Hawthorn_2019}. Each model runs for 3.6 Myr during which the flare is on for 0.6 Myr and then is instantaneously turned off. We do not include models where the flare decays over time, because a reasonable decay time has been shown to result in only small changes when compared to an instantaneous turnoff in Magellanic stream models\cite{Bland_Hawthorn_2013}.
The combined Galactic and extragalactic radiation fields, along with the cosmic ray background are kept constant for the duration of the calculation. The models use a self-consistent approach in which the cloud is assumed to be in ionization equilibrium before the flare event occurs, thus we adopt as starting conditions the observed ion abundance, log $n_\mathrm{H}$, and cloud size (=$N_\mathrm{H}$/$n_\mathrm{H}$) from the ionization equilibrium model. After each time step of 0.2 Myr, the relative ionization fraction correction is calculated for the species of interest. A summary of these results is given in Extended Data Fig.~\ref{table:cloudy}.

Extended Data Fig.~\ref{fig:IC} shows the time-dependent \texttt{CLOUDY} results from the beginning of the Seyfert flare event ($t$=0 Myr) to current day (3.6 Myr). Since there is a large change in IC immediately after the Seyfert flare is shut off, followed by a period of relatively small variations, we plot a magnified version of the IC from 0.8--3.6 Myr as an inset panel to show these small fluctuations. After the flare is shut off (0.6 Myr), both the ICs and column densities begin to rapidly change by 0.8 Myr and stabilize by $\sim$1 Myr for all sight lines. Both sight lines approach the ICs of the equilibrium models well before 3.6 Myr. Therefore, for a Seyfert flare occurring 3.6 Myr ago with a log\,$\Phi$(H)=8--10 and without a long fading period, the FB HVCs have had sufficient time to reach an equilibrium ionization state and their ICs are largely unchanged by the flare.

We calculate the FB HVCs' metallicity by adding the IC calculated from log\,$\Phi$(H)=10 to the ion abundances calculated from Equation~\ref{equation:XiH}, estimating the error in IC as the difference between the ICs calculated from log\,$\Phi$(H) of 8 and 10. The results are shown in Extended Data Fig.~\ref{table:metallicities}.

J1919-2958's and 1H1613-097's metallicity measurements made using the time-dependent ICs in Extended Data Fig.~\ref{table:metallicities} have limits on their ICs due to saturation of Si\;{\small{\MakeUppercase{\romannumeral 3}}}\ 1206 absorption, leading to lower limits on the Si\;{\small{\MakeUppercase{\romannumeral 3}}}\ to Si\;{\small{\MakeUppercase{\romannumeral 2}}}\ ratio. This translates to upper limits on the gas density. In Supplementary Information Section 5, we show that a 1 dex difference in Si\;{\small{\MakeUppercase{\romannumeral 3}}}\ columns does not strongly affect the ICs or leads to nonphysical models. We therefore have removed the lower limits in Table~\ref{table:main_metallicities} and Extended Data Fig.~\ref{table:metallicities} for the metallicity measurements.

\subsection*{Literature Sample Ionization Corrections}

We do not include LS 4825-1 or LS 4825-2 in the \texttt{CLOUDY} modeling described above because neither have published Si\;{\small{\MakeUppercase{\romannumeral 2}}}\ or Si\;{\small{\MakeUppercase{\romannumeral 3}}}\ measurements.
Therefore, the clouds' gas density, derived from Si\;{\small{\MakeUppercase{\romannumeral 3}}}/Si\;{\small{\MakeUppercase{\romannumeral 2}}}, cannot be constrained. We instead use the grid of \texttt{CLOUDY} equilibrium models previously calculated in the literature for 1H1613-097 to estimate the IC for LS 4825-1's and LS 4825-2's FB HVCs\cite{Bordoloi_2017}.  We determine the ICs for the FB HVCs using the measured $N_{\mathrm{HI}}$ and a range of $\mathrm{log}~U$ values ($-3.5$, $-3.0$, and $-2.5$), which reflect the values derived for other Milky Way halo and Fermi Bubble clouds in the literature\cite{Collins_2005, Richter_2009, Tripp_2011, Fox_2014, Fox_2016, Bordoloi_2017}. For the metallicity calculations (see Extended Data Fig.~\ref{table:metallicities}) we use the IC value from $\mathrm{log}~U=-3.0$ and the other two IC values ($\mathrm{log}~U=-3.5$ and $-2.5$) serve as the errors on the IC. 

We also use the equilibrium \texttt{CLOUDY} models previously calculated for the 1H1613-097 sight line\cite{Bordoloi_2017} to estimate the ionization corrections for FB HVCs with O\;{\small{\MakeUppercase{\romannumeral 1}}}\ detections and H\;{\small{\MakeUppercase{\romannumeral 1}}}\ limits. For this sample we use the IC values calculated from $\mathrm{log}~U=-3.5$ for the H\;{\small{\MakeUppercase{\romannumeral 1}}}\ column limit as it provides an IC(O) lower limit to add to our lower-limit metallicity calculations.

We did not calculate an IC for PKS 2005-489. PKS 2005-489's metallicity estimate, drawn from the literature, was made by comparing the FB HVC's FUSE and STIS absorption line detections (Si\;{\small{\MakeUppercase{\romannumeral 2}}}, C\;{\small{\MakeUppercase{\romannumeral 3}}}, O\;{\small{\MakeUppercase{\romannumeral 6}}}, and H\;{\small{\MakeUppercase{\romannumeral 1}}}\ Lyman series) to those of highly ionized Milky Way halo clouds with similar absorption columns and existing \texttt{CLOUDY} models\cite{Collins_2004, Keeney_2006}. 

The IC for M5-ZNG1 is calculated with the \texttt{CLOUDY} equilibrium modeling described above. We chose an equilibrium model because the Seyfert flare event 3.5 Myr ago was shown not to have a lasting effect on the present-day ICs in other FB HVCs. Therefore, a full time-dependent treatment was deemed unnecessary. The equilibrium IC is shown in Extended Data Fig.~\ref{table:metallicities}.

%TC:ignore

\section*{Data availability}
The HST/COS datasets for all sources used in this paper are available in MAST with the identifier \url{http://dx.doi.org/10.17909/zxzh-4x54}\cite{Ashley_2022}, including HST Program IDs 1533, 7345, 8096, 9410, 11741, 12603, 13448, and 15339, and FUSE Program IDs A108, C149, D157, and P107. The GBT raw datasets are publicly available at the NRAO archive, \url{https://data.nrao.edu}; GBT programs for all sources in this paper with GBT data are GBT14B-299, GBT15B-359, GBT16B-422, GBT17B-015, GBT18A-221, GBT20A-253, and 
GBT20B-444. Data from GBT20B-444 will become available on the NRAO archive after the proprietary period ends July 9, 2022. Green Bank 140' data used for calculating PG 1522+101's HI column density limit can be requested through \url{https://help.nrao.edu/}. LAB data used for calculating J1509-0702's HI column density limit is publicly available at \url{https://www.astro.uni-bonn.de/hisurvey/AllSky_profiles}. Fully reduced data are available from the corresponding author on reasonable request.

\section*{Code availability} 

The IDL software used in this work is available for purchase from \url{https://www.l3harrisgeospatial.com/Software-Technology/IDL}.  The GBTIDL, VPFIT, and CLOUDY software are publicly available. The GBTIDL package for GBT data reduction and analysis can be downloaded from \url{https://gbtidl.nrao.edu/downloads.shtml}. The VPFIT package for fitting Voigt profiles to absorption data can be found at \url{https://people.ast.cam.ac.uk/~rfc/}. The CLOUDY cloud-modeling code is available at \url{https://gitlab.nublado.org/cloudy/cloudy/-/wikis/DownloadLinks}.

\section*{Acknowledgements}

We would like to thank Joss Bland-Hawthorn for his valuable discussion on ionization from the Seyfert flare event. Support for T.A. through \emph{HST} program 15339 was provided by NASA through grants from the Space Telescope Science Institute, which is operated by the Association of Universities for Research in Astronomy, Inc., under NASA contract NAS 5-26555. The Green Bank Observatory is a facility of the National Science Foundation, operated under a cooperative agreement by Associated Universities, Inc. The GBT data presented in this paper were obtained under Programs GBT14B-299, GBT15B-359, GBT16B-422, GBT17B-015, GBT18A-221, GBT20A-253, and GBT20B-444.

\section*{Author Contributions}
\noindent T.A. led the investigation, including the sample curation, UV and radio measurements, analysis, and writing.
A. J. F. led the project conceptualization and management, is the PI of the HST program that funded the research, and contributed directly to the writing of the paper.
F.H.C. performed the CLOUDY modeling, contributed directly to the writing of the paper, and prepared Extended Data Fig.~\ref{fig:IC}.
F.J.L. reduced and analyzed the GBT data and contributed directly to the writing of the paper.
R.B. provided the 1H1613-097 GBT data and its data reduction.
B.P.W. provided the data reduction for the UV data. 
E.B.J. provided an interpretation of the cloud destruction and survival. 
T.K. prepared Figure~2 and Supplementary Figure~2.   
All authors reviewed the manuscript, contributed to the editing of the manuscript, and contributed to the discussion of the results' interpretation and implications.

\subsection*{Corresponding author}
\noindent Correspondence to Trisha Ashley.
\href{mailto:tashley@stsci.edu}{\Letter}

\section*{Competing interests}
The authors declare no competing interests.

\clearpage

\section*{Extended Data Figures}

\setcounter{figure}{0}

\begin{figure}[!ht]
    \renewcommand{\figurename}{Extended Data Figure} 
   \begin{center}
     \includegraphics[ width=\textwidth]{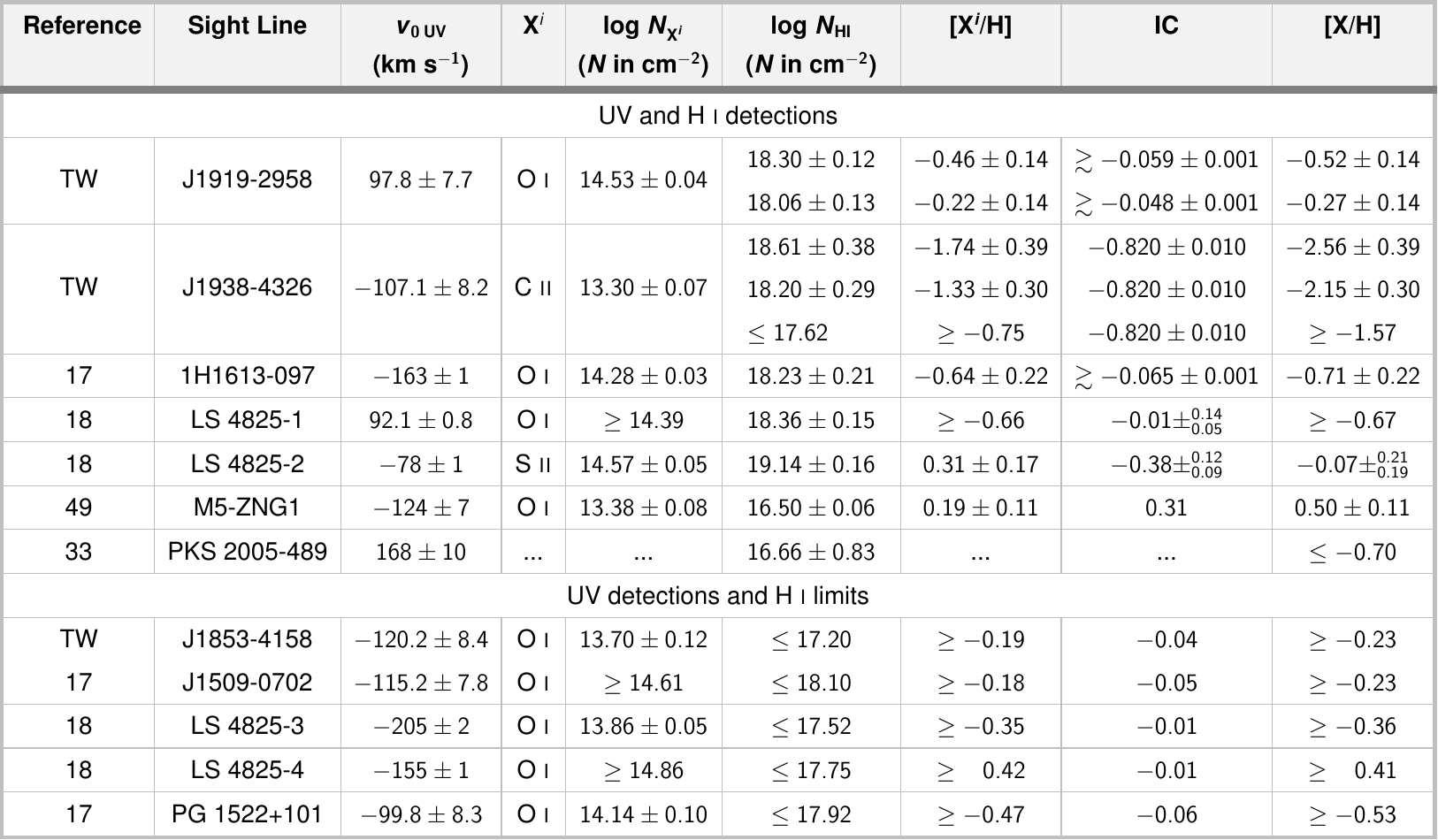}
    \end{center}
\caption{Full Metallicity Measurements. $X^{i}$ represents the UV-detected ion used for metallicity calculations and $v_{0\mathrm{UV}}$ is the FB HVCs' LSR velocity centroid for that ion. The ion absorption and HI emission, log $N_{\mathrm{X}^i}$. and $N_{\mathrm{HI}}$, respectively, are listed with the HI column errors including beam smearing added in quadrature when available. We also list the ion abundances, [X$^{i}$/H]. The ionization correction, IC, calculations are discussed throughout the Methods section. [X/H] is the corrected gas-phase elemental abundance and does not account for dust depletion. We include an OI solar abundance for M5-ZNG1's elemental abundance measurement that is updated from that in the literature\cite{Keeney_2006, Asplund_2009}. For a discussion of the multiple HI measurements for J1919-2958, see the Methods Section. For a discussion of dust depletion and the multiple H\;{\small{\MakeUppercase{\romannumeral 1}}}\ measurements for J1938-4326, see Supplementary Information Section 1.}\label{table:metallicities}
\end{figure}

\begin{figure}[!ht]
    \renewcommand{\figurename}{Extended Data Figure} 
   \begin{center}
     \includegraphics[ width=\textwidth]{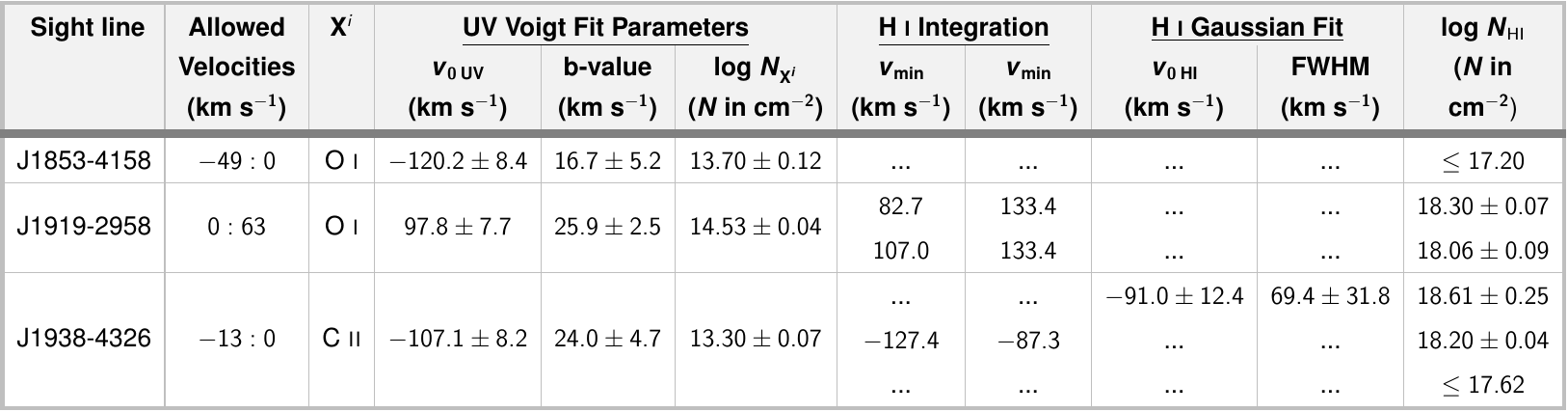}
    \end{center}
\caption{UV and HI Fit Parameters for New Sight Lines. The second column represents the allowed velocity range of gas co-rotating with the Milky Way disk in each quasar's direction\cite{Wakker_1991}. X$^{i}$ represents the ion used for metallicity calculations. The UV Voigt fit parameters of X$^{i}$ for each cloud are the LSR velocity centroid, $v_{0\:\mathrm{UV}}$, the Doppler broadening parameter (b-value), and the log column density, log($N_{\mathrm{X}^i}$). The UV velocity centroid errors include the 7.5 km~s$^{-1}$\  COS zero-point uncertainty. The H\;{\small{\MakeUppercase{\romannumeral 1}}}\  Gaussian fit parameters are the LSR velocity centroid, $v_{0\:\mathrm{HI}}$, full-width-half-max, FWHM, and log column density, log $N_{\mathrm{H\;I}}$. For a Gaussian profile, the relation between FWHM and $b$-value is FWHM=1.665$b$. J1853-4158's H\;{\small{\MakeUppercase{\romannumeral 1}}}\  column was obtained using the spectrum's rms, as described in the Methods Section. J1919-2958's H\;{\small{\MakeUppercase{\romannumeral 1}}}\  column was obtained by through the ``flip-and-subtract'' method (described in the Methods Section) using two velocity ranges for integration, which encompass all potential emission associated with the FB HVC (upper column limit) and emission least affected by stray radiation (lower column limit). J1938-4326's H\;{\small{\MakeUppercase{\romannumeral 1}}}\ column was obtained using a Gaussian fit to emission. Second and third measurements of the H\;{\small{\MakeUppercase{\romannumeral 1}}}\ column were made by integrating over the C\;{\small{\MakeUppercase{\romannumeral 2}}}\ absorber's FWHM and then using Equation~\ref{Equation:HI_limit}; see Supplementary Information Section 1 for more details.}\label{table:UV_HI_fits}
\end{figure}

\begin{figure}[!ht]
    \renewcommand{\figurename}{Extended Data Figure} 
   \begin{center}
     \includegraphics[ width=\textwidth]{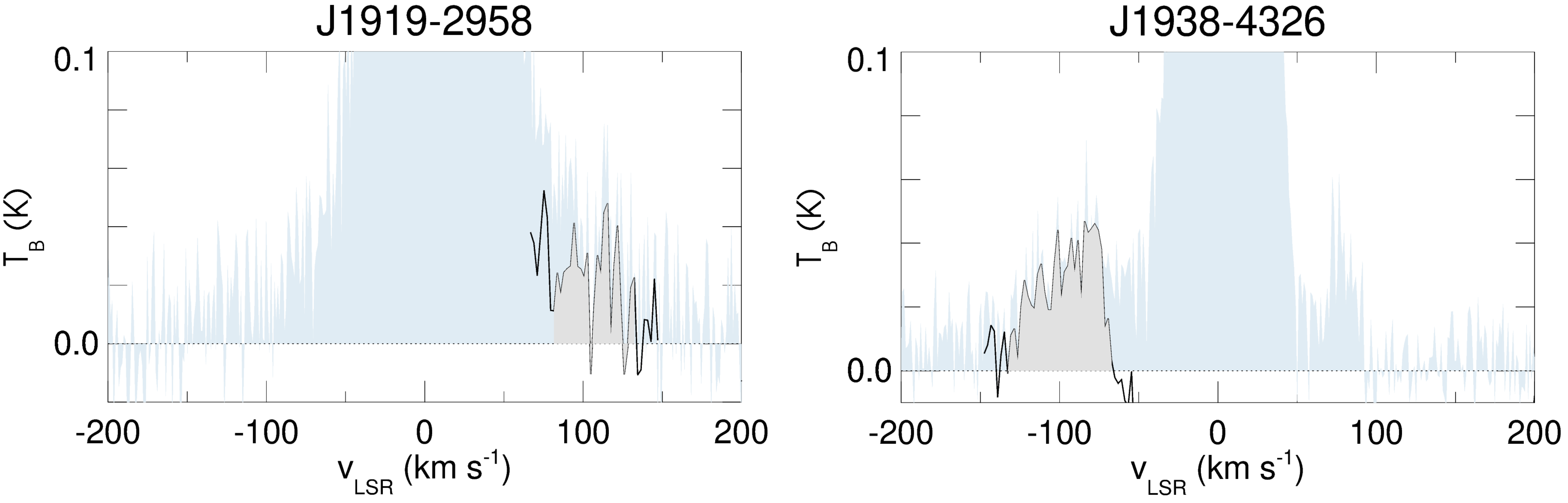}
    \end{center}
\caption{Flipped-and-subtracted GBT H\;{\small{\MakeUppercase{\romannumeral 1}}}\ spectrum for J1919-2958 and 1938-4326. The blue shaded spectrum represents the original H\;{\small{\MakeUppercase{\romannumeral 1}}}\ spectrum at a resolution of $\sim$1 km~s$^{-1}$. The maximum velocity range used to calculate the H\;{\small{\MakeUppercase{\romannumeral 1}}}\  column densities is shaded in grey. The black line is the flipped-and-subtracted spectra smoothed to  $\sim$2 km~s$^{-1}$\  in the integrated velocity ranges including an additional 7 channels on each side of those velocity ranges. }\label{figure:1919_hi_flip} 
\end{figure}

\begin{figure}[!ht]
    \renewcommand{\figurename}{Extended Data Figure} 
   \begin{center}
     \includegraphics[ width=\textwidth]{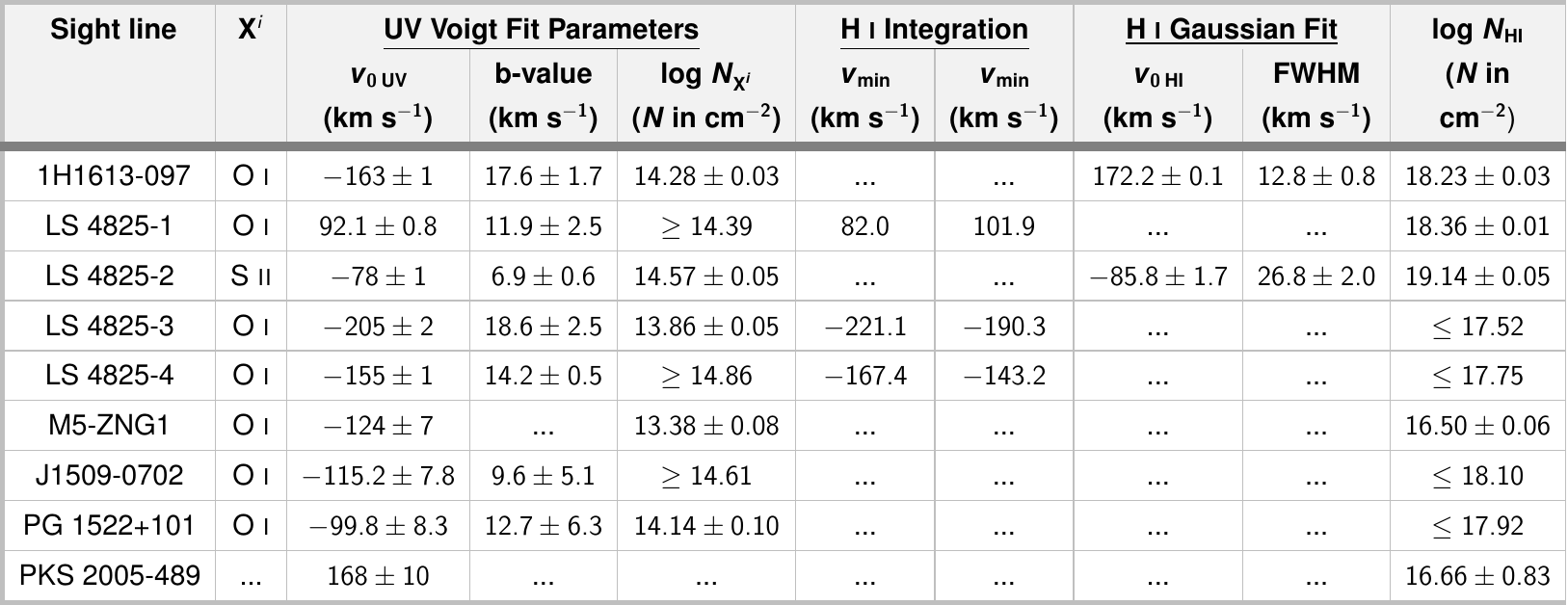}
    \end{center}
\caption{UV and H\;{\small{\MakeUppercase{\romannumeral 1}}}\ Fit Parameters for Literature Sight Lines. This table is lists UV and H\;{\small{\MakeUppercase{\romannumeral 1}}}\  fit parameters for the literature FB HVCs, similar to that done for the new sight lines in Supplementary Table~\ref{table:UV_HI_fits}. M5-ZNG1 has a literature H\;{\small{\MakeUppercase{\romannumeral 1}}}\  measurement based on a combination of FUSE profile fitting and curve-of-growth measurements\cite{Zech_2008}. J1509-0702 and PG 1522+10 have H\;{\small{\MakeUppercase{\romannumeral 1}}}\   measurements based on their average rms of emission-free channels (rms of 0.055 and 0.032 K, respectively). PKS 2005-489's log $N_\mathrm{HI}$ is from Lyman series measurements in the literature\cite{Keeney_2006}.}\label{table:UV_HI_fits_literature}
\end{figure}

\begin{figure*}[!t]
 \renewcommand{\figurename}{Extended Figure} 
    \centering
 \includegraphics[width=\textwidth]{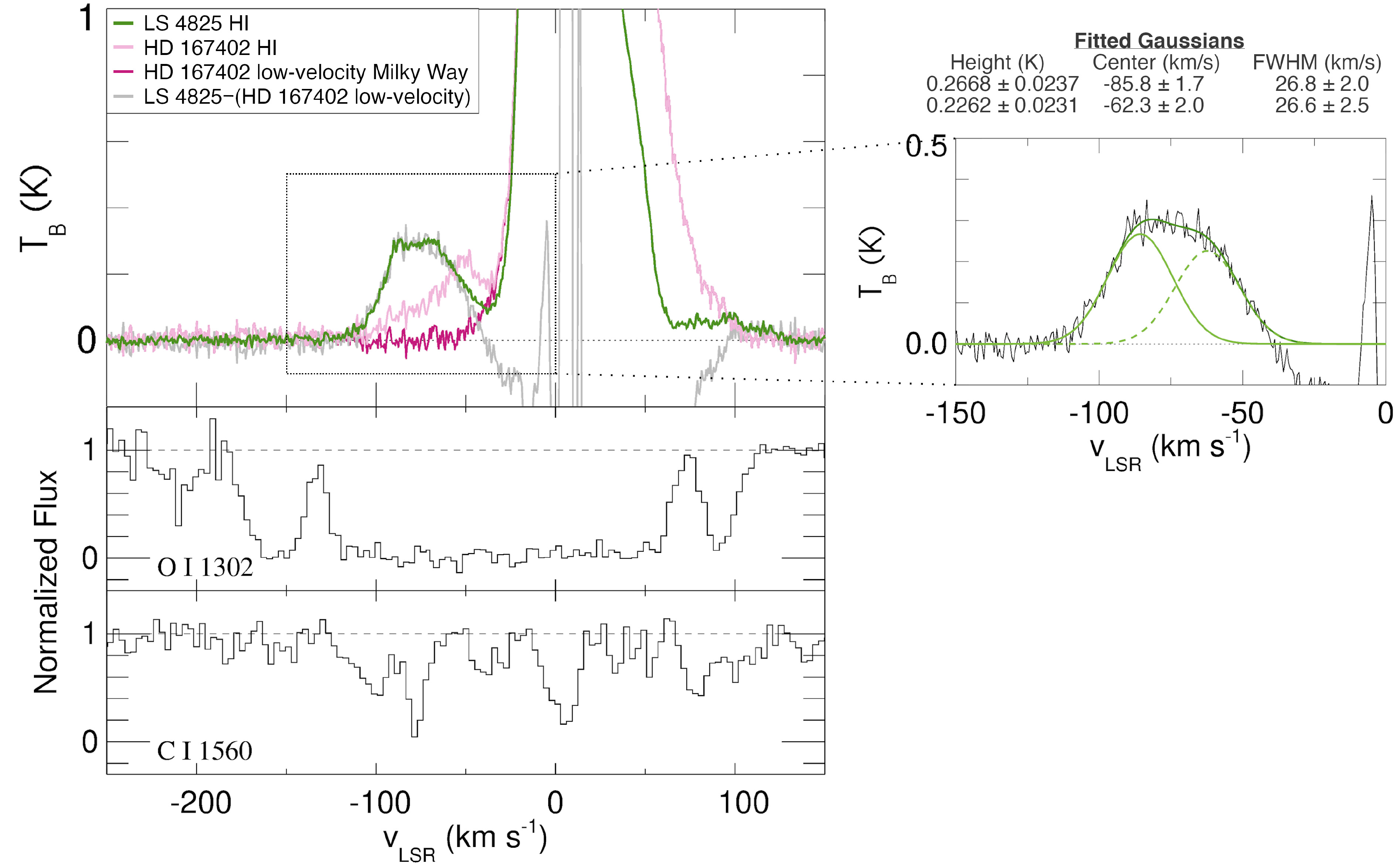}
\caption{Deconvolution of LS 4825's H\;{\small{\MakeUppercase{\romannumeral 1}}}\ spectrum. Left: LS 4825's and HD 167402's H\;{\small{\MakeUppercase{\romannumeral 1}}}\ spectra plotted against HD 167402's low-velocity Milky Way component and LS 4825's residual \H\;{\small{\MakeUppercase{\romannumeral 1}}}\ spectrum after subtracting HD 167402's low-velocity Milky Way component. We also plot LS4825's O\;{\small{\MakeUppercase{\romannumeral 1}}}\ $\lambda$1302 and C\;{\small{\MakeUppercase{\romannumeral 1}}}\  $\lambda$1560 spectrum for comparison. Right: The individual and combined Gaussian fits to LS 4825's residual negative-velocity components. }\label{figure:LS4825_HI} 
\end{figure*}

\begin{figure}[!ht]
    \renewcommand{\figurename}{Extended Data Figure} 
   \begin{center}
     \includegraphics[ width=\textwidth]{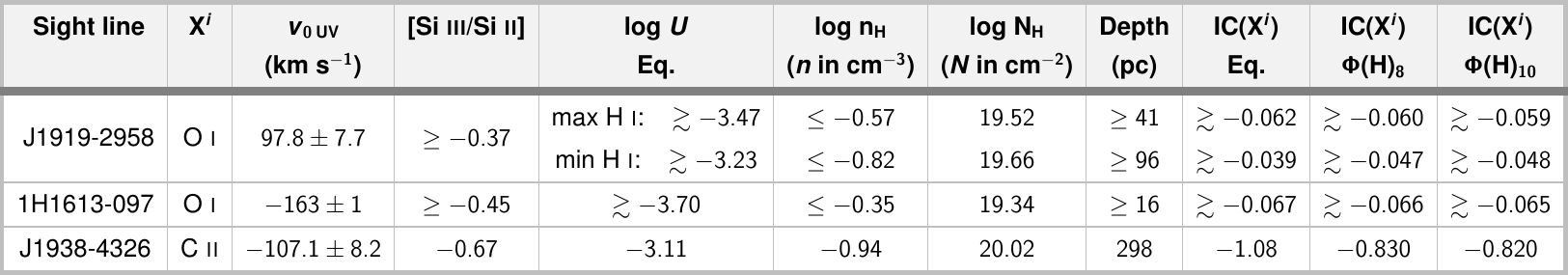}
    \end{center}
\caption{X$^{i}$ represents the ion used for the metallicity calculations and $v_{0\:\mathrm{UV}}$ is the UV velocity centroid of each component. The logarithmic Si\;{\small{\MakeUppercase{\romannumeral 3}}}\  to Si\;{\small{\MakeUppercase{\romannumeral 2}}}\  ion ratio, [ Si\;{\small{\MakeUppercase{\romannumeral 3}}}\  /Si\;{\small{\MakeUppercase{\romannumeral 2}}}, is calculated using the column densities measured in previous Fermi Bubble UV Surveys\cite{Bordoloi_2017, Ashley_2020}; for J1938-4326 we use the [Si\;{\small{\MakeUppercase{\romannumeral 3}}}\ /C\;{\small{\MakeUppercase{\romannumeral 2}}}] ratio. 
$U$ is the ionization parameter, equal to the ratio of the ionizing photon density to the gas density. The log hydrogen number density and column density of the clouds are given as log $n_\mathrm{H}$ and log $N_\mathrm{H}$, respectively. The depth is the calculated size of the cloud along the line-of-sight. The present day ionization corrections are given as IC(X$^{i}$) Eq. for the equilibrium models, and IC(X$^{i}$) $\Phi(\mathrm{H})_{8}$ and IC(X$^{i}$) $\Phi(\mathrm{H})_{10}$ for the time-dependent models at two ionizing photon fluxes, log\,$\Phi$(H)=8 and 10, respectively.}\label{table:cloudy}
\end{figure}

\begin{figure}[!ht]
    \renewcommand{\figurename}{Extended Data Figure} 
   \begin{center}
     \includegraphics[ width=\textwidth]{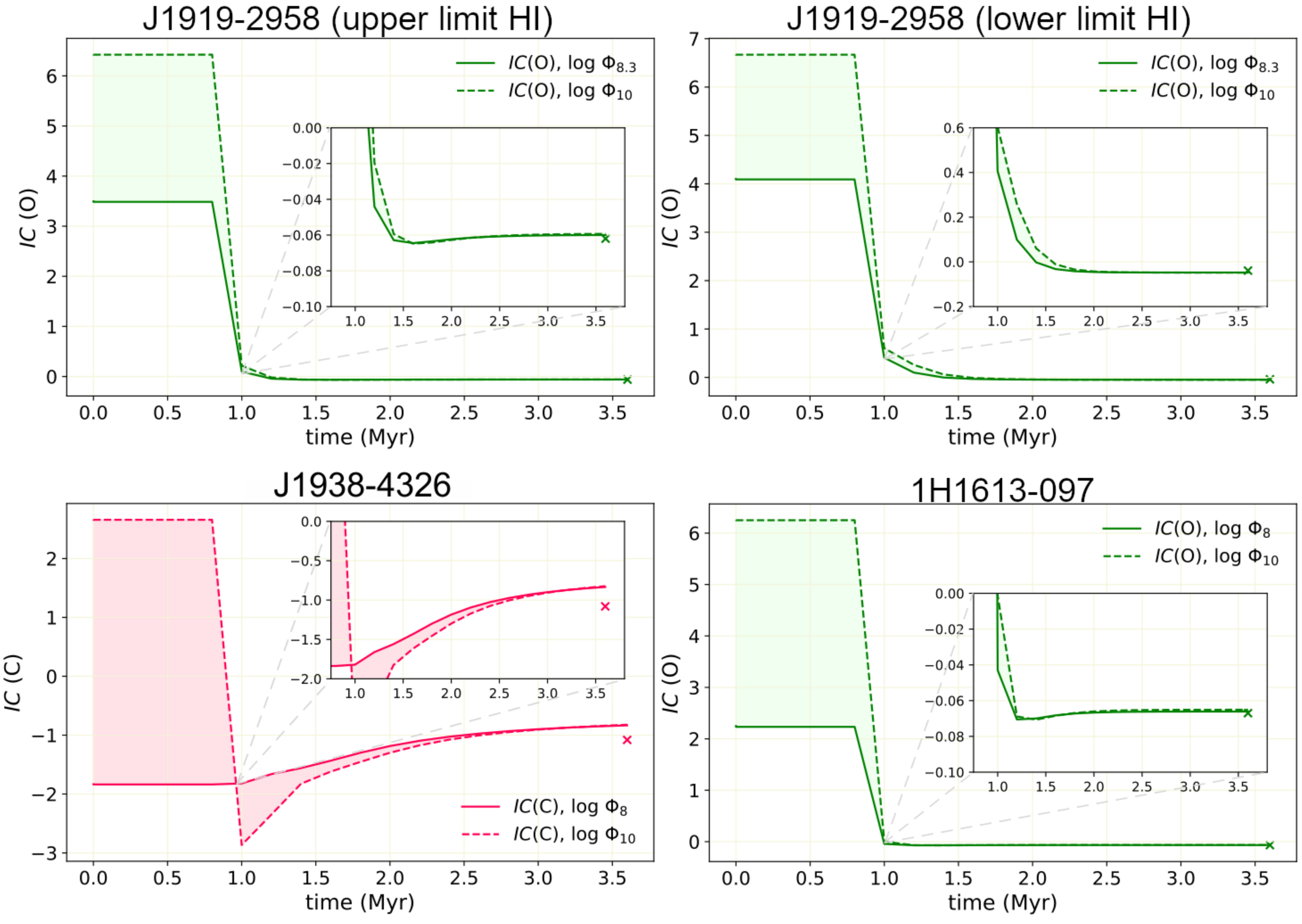}
    \end{center}
\caption{Time-dependent ionization corrections versus time calculated from our \texttt{CLOUDY} models. The results are given for two ionizing fluxes: log $\Phi(\mathrm{H})$ = 8 (solid line) and 10 (dashed line), where $\Phi$(H) has units of photons cm$^{-2}$ s$^{-1}$ and where colored shading shows intermediate ionizing fluxes. 
Each panel shows the ionization correction across the full time interval modeled (0--3.6 Myr), with an inset plot magnifying the flatter part of the curves after the initial flash to emphasize the late-time behavior. The equilibrium results are marked with an `x' at 3.6 Myr in each panel.}\label{fig:IC}
\end{figure}

\clearpage

\bibliographystyle{naturemag}

\bibliography{Metallicity_Galactic_Center_4AGN}

\clearpage
\section*{Supplementary Information to Diverse Metallicities of Fermi Bubble Clouds Indicate Dual Origins in the Disk and Halo}

%TC:ignore

\section{Dust depletion and J1938-4326's low metallicity}\vspace{12pt}

Dust depletion does not strongly affect oxygen measurements and is largely thought to not strongly affect sulfur measurements\cite{Savage_1996, Jenkins_2009}. However, dust depletion can be significant for carbon, which is used to measure the metallicity for J1938-4326. Without taking into account dust depletion affects, the log of J1938-4326's gas-phase elemental abundance is 0.28\%\ solar (see Extended Data Fig.~\ref{table:metallicities}). In the densest clouds, carbon levels can be depleted by up to $-0.58$ dex \cite{Jenkins_2009}\textsuperscript{,68}. That would raise the metallicity of the FB HVC towards J1938-4326 to $1$\%\ solar (or $-1.98$ dex in log space), a resolutely low metallicity.

With such a low metallicity, we also explore the possibility that the \HI\ spectrum towards J1938-4326 may have emission from multiple clouds contributing to the \HI\ Gaussian fit shown in Figure~\ref{figure:gauss_fit}, which may not all be associated with the FB HVC towards J1938-4326. To do this, we integrate the \HI\ emission over the velocity range of the \CII\ absorption's FWHM and centered on the \CII\ absorption's $v_{0\:\mathrm{UV}}$ ($-127$ to $-87$ \kms). This results in a log $N_{\mathrm{HI}}=18.20$. Assuming the same IC as listed in Extended Data Fig.~\ref{table:metallicities} and a dust depletion of $-0.58$ dex, the resulting metallicity would still only reach $2.7$\%\ solar. If we instead assume that none of the \HI\ detected in this velocity range is associated with the FB HVC and use Equation~\ref{Equation:HI_limit} to estimate an upper limit on the \HI\ column density, then we measure log $N_{\mathrm{HI}}\le17.62$; assuming the same IC and dust depletion, the metallicity is raised to only $10$\%\ solar. Therefore, the data indicate that this FB HVC has a low metallicity even when dust depletion is taken into account. In Table~\ref{table:main_metallicities} we include J1938-4326's metallicity as a range between that which is calculated using the Gaussian fit to the \HI\ and that which is  calculated using Equation~\ref{Equation:HI_limit}; both limits contain a dust depletion level of $-0.58$ dex. We also include the various \HI\ column measurements in Extended Data Fig.~\ref{table:metallicities} and \ref{table:UV_HI_fits}.\looseness=-2

\section{Check of J1938-4326's HI Gaussian fit}\vspace{12pt}

To verify the Gaussian-fit \HI\ column density measurement of J1938-4326 in Extended Data Fig.~\ref{table:UV_HI_fits}, we use the same flip-and-subtract measurement method that is used to determine J1919-2958's \HI\ column. The remaining ``flipped-and-subtracted" spectra is shown in Extended Data Fig.~\ref{figure:1919_hi_flip}. We integrate the emission from $-131.7$ to $-67.2$ km~s$^{-1}$\ and calculate a log~$N_{\mathrm{HI}}$ of $18.53\pm0.04$ using Equation~\ref{equation:nhi}; this log~$N_{\mathrm{HI}}$ from the flip-and-subtract method is within the errors of that listed in  Extended Data Fig.~\ref{table:UV_HI_fits}. If we instead use the FWHM and velocity centroid of J1938-4326's C\;{\small{\MakeUppercase{\romannumeral 2}}}\ absorption to integrate over the residual spectrum ($-127.4$ to $-87.3$ km~s$^{-1}$), then the log~$N_{\mathrm{HI}}$ decreases by 0.21 dex or 0.04 dex below that of column minus the error listed in  Extended Data Fig.~\ref{table:UV_HI_fits}. 

\section{Effects of the XX-polarization on GBT maps} 
All XX-polarization data has been discarded in the GBT \HI\ emission maps presented in Figure~\ref{figure:hi_maps} due to spurious features at levels of tens of mK produced by frequency switching. Typically these features can be ignored due to their low levels; however, \HI\ emission of the FB HVCs in this work also appear on the level of 10-20 mK (see Supplementary Fig.~\ref{figure:xx_pol}), making it possible to misinterpret these spurious features as emission. The YY polarization only suffers minor spurious waves that are not likely to be misinterpreted as emission. An example of the difference between the YY-only and XX-YY combined spectra from the GBT maps is shown in Supplementary Fig.~\ref{figure:xx_pol}. In the single-pointing H\;{\small{\MakeUppercase{\romannumeral 1}}}\ data near the emission velocity regions these spurious features were not as pronounced; therefore, we left the XX-polarization in the single-pointing data to obtain the better signal-to-noise for the column density measurements presented in Extended Data Fig.~\ref{table:UV_HI_fits}.

\section{\texttt{CLOUDY} model comparison for J1919-2958's range of HI columns}
Because of the range in J1919-2958's H\;{\small{\MakeUppercase{\romannumeral 1}}}\ column density, we run two sets of \texttt{CLOUDY} models for this sight line. \texttt{CLOUDY} produces a list of predicted ion columns for each model. In the equilibrium models of both the lower and upper H\;{\small{\MakeUppercase{\romannumeral 1}}}\ column limit, we compared the FB HVC's measured ion columns to \texttt{CLOUDY}'s predicted columns (for Al\;{\small{\MakeUppercase{\romannumeral 2}}}\ $\lambda$1670, C\;{\small{\MakeUppercase{\romannumeral 2}}}\ $\lambda$1334, C\;{\small{\MakeUppercase{\romannumeral 4}}}\ $\lambda$1548, 1550, Fe\;{\small{\MakeUppercase{\romannumeral 2}}}\ $\lambda$1144, 1608, O\;{\small{\MakeUppercase{\romannumeral 1}}}\ $\lambda$1302, S\;{\small{\MakeUppercase{\romannumeral 2}}}\ $\lambda$1250, 1253, 1259, Si\;{\small{\MakeUppercase{\romannumeral 2}}}\ $\lambda$1190, 1193, 1260, 1526, Si\;{\small{\MakeUppercase{\romannumeral 3}}}\ $\lambda$1206, and Si\;{\small{\MakeUppercase{\romannumeral 4}}}\ $\lambda$1393, 1402). We find that of the two \texttt{CLOUDY} models, the upper H\;{\small{\MakeUppercase{\romannumeral 1}}}\ column density limit model more closely coincides with the measured column densities. The average absolute relative errors in the linear column densities are 2.2 and 0.87 for the lower and upper H\;{\small{\MakeUppercase{\romannumeral 1}}}\ column density limits, respectively. Therefore, the \texttt{CLOUDY} models slightly favor the upper H\;{\small{\MakeUppercase{\romannumeral 1}}}\ column density limit and lower metallicity for the FB HVC toward J1919-2958.

\section{Ionization correction dependence on saturation of Si~III }
J1919-2958's and 1H1613-097's ICs in Extended Data Fig.~\ref{table:metallicities} are limits due to saturation of Si\;{\small{\MakeUppercase{\romannumeral 3}}}\ 1206 absorption, resulting in upper limits on the gas density. However, IC(O) values tend to be near zero for log~$N_{\mathrm{HI}}$\textgreater 18.5$^[$\cite{Bordoloi_2017}$^]$. We therefore test the robustness of these metallicity measurements by changing the saturated log~$N_{\mathrm{Si\;III}}$ measurement by 1~dex and recalculating the metallicities using the ionization equilibrium models. We show the model parameters and results in Supplementary Table~\ref{table:metallicity_robustness}. The metallicity measurements for 1H1613-097 and J1919-2958's upper H\;{\small{\MakeUppercase{\romannumeral 1}}}\ column density limit from a 1 dex increase in Si\;{\small{\MakeUppercase{\romannumeral 3}}}\ columns are approximately within the metallicity errors listed in Extended Data Fig.~\ref{table:metallicities}. This 1 dex increase shows that the metallicity results are not strongly dependent on Si\;{\small{\MakeUppercase{\romannumeral 3}}}\ saturation for high H\;{\small{\MakeUppercase{\romannumeral 1}}}\ columns. Conversely, J1919-2958's lower H\;{\small{\MakeUppercase{\romannumeral 1}}}\ column density limit (log~$N_{\mathrm{HI}}$=18.06) does have a significant change in IC of about $-0.41$ dex when the Si\;{\small{\MakeUppercase{\romannumeral 3}}}\ measurement increases by 1 dex. We note, however, that J1919-2958's lower limit H\;{\small{\MakeUppercase{\romannumeral 1}}}\ column model results for the increase in 1 dex give an unrealistic cloud depth of $\sim$6 kpc. Additionally, the \texttt{CLOUDY} model with log~$N_{\mathrm{Si\;III}}$=13.71 has predicted ion columns significantly closer to the measured columns, with an average absolute relative error on the linear ion columns of 2.2, whereas the model with log~$N_{\mathrm{Si\;III}}$=14.71 has an average absolute relative error of 25. We therefore remove the metallicity measurements' lower limits in Table~\ref{table:main_metallicities} and Extended Data Fig.~\ref{table:metallicities}.

\clearpage

\section{Supplementary tables}\vspace{12pt}

\setcounter{figure}{0}    

\begin{figure*}[!ht]
 \renewcommand{\figurename}{Supplementary Table} 
 \begin{center}
      \includegraphics[ width=0.5\textwidth]{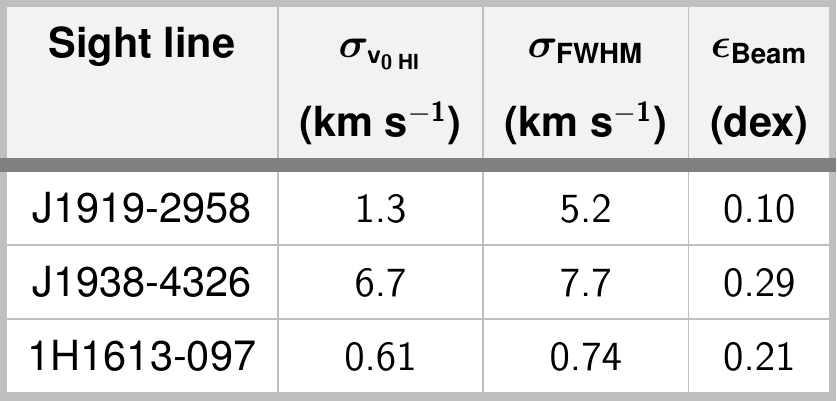}
  \end{center}
\caption{Beam Smearing Estimates. This table quantifies the beam smearing uncertainties of our \HI\ measurements. The standard deviations of the \HI\ velocity centroid and FWHM are given as $\sigma_{v_{0\:\mathrm{HI}}}$ and $\sigma_{\mathrm{FWHM}}$, respectively, measured by comparing pointings along the central quasar sight line with those centered a half beamwidth away. $\epsilon_{\mathrm{Beam}}$ is the beam-smearing error on the \HI\ column density.} \label{table:beam_smearing}
\end{figure*}

\begin{figure*}[!ht]
 \renewcommand{\figurename}{Supplementary Table} 
      \includegraphics[ width=\textwidth]{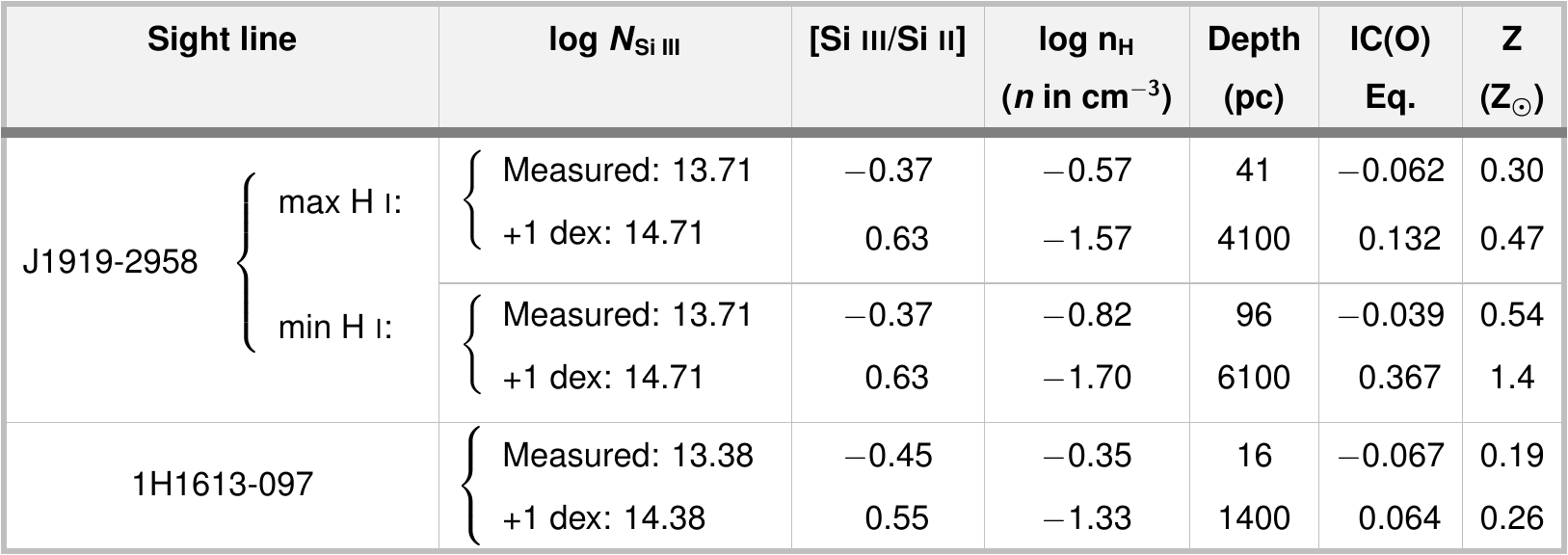}
\caption{For each absorber we calculate the effect on IC(O) from a 1 dex increase in log\,$N_\mathrm{Si\;III}$. This table demonstrates the marginal effect of different degrees of saturation in \ion{Si}{iii} 1206 on most oxygen ionization corrections. Only the minimum \HI\ measurements for J1919-2958 show a significant change in metallicity; however, the  model produces unrealistic cloud depths and ion columns that have high average absolute relative errors on the ion column densities.  
}\label{table:metallicity_robustness}
\end{figure*}

\clearpage
\section{Supplementary figures}\vspace{12pt}

\setcounter{figure}{0}

\begin{figure*}[!ht]
 \renewcommand{\figurename}{Supplementary Figure} 
      \includegraphics[width=\textwidth]{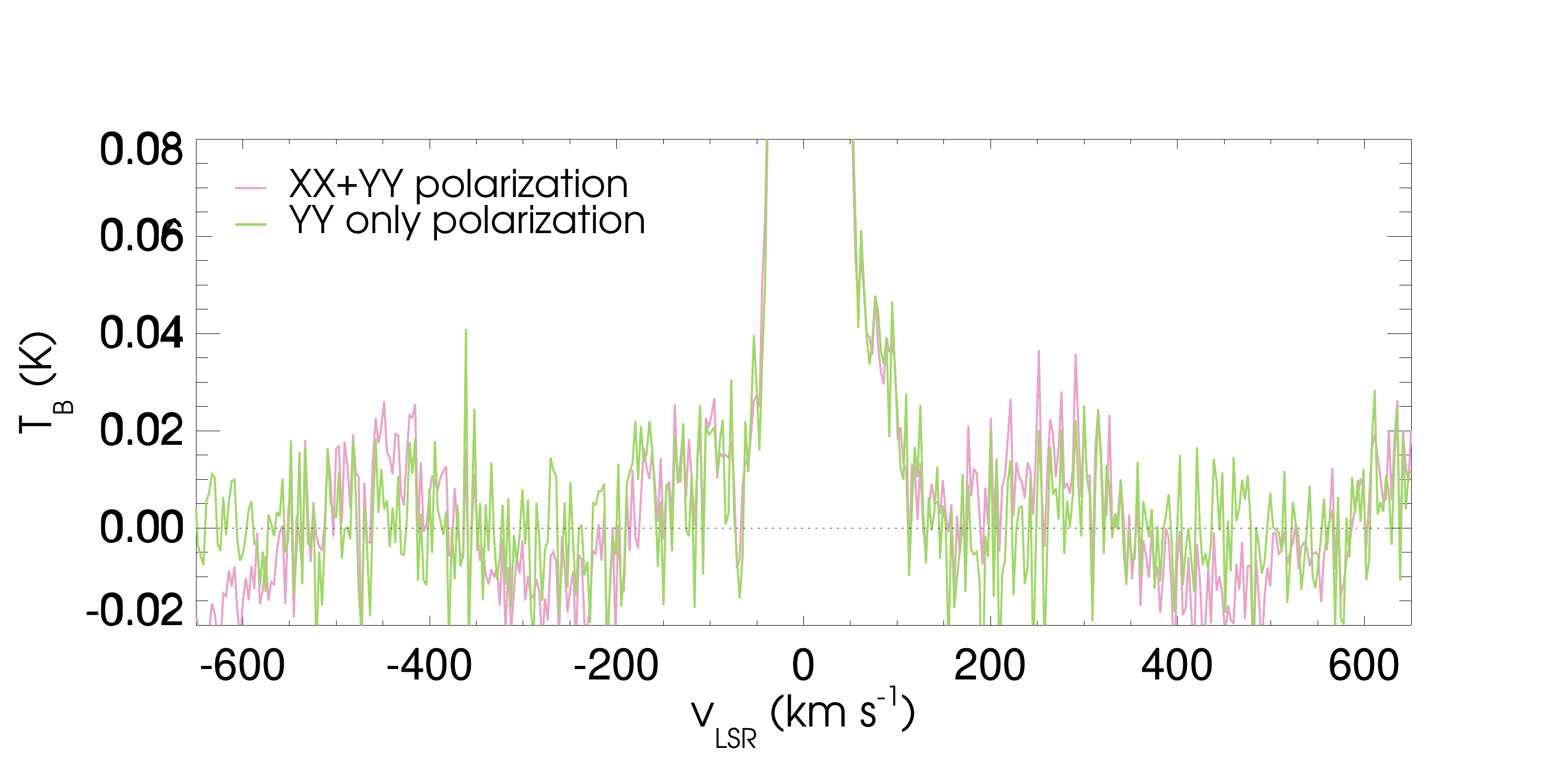}
\caption{Spectra from J1919-2958 GBT \HI\ maps comparing both polarizations combined (XX+YY; pink) and only the YY-polarization (green). This figure demonstrates that the spurious spectral features in the maps are prominent in the XX-polarization.}\label{figure:xx_pol}
\end{figure*}

\begin{figure*}[!ht]
 \renewcommand{\figurename}{Supplementary Figure} 
    \begin{center}
      \includegraphics[ width=\textwidth]{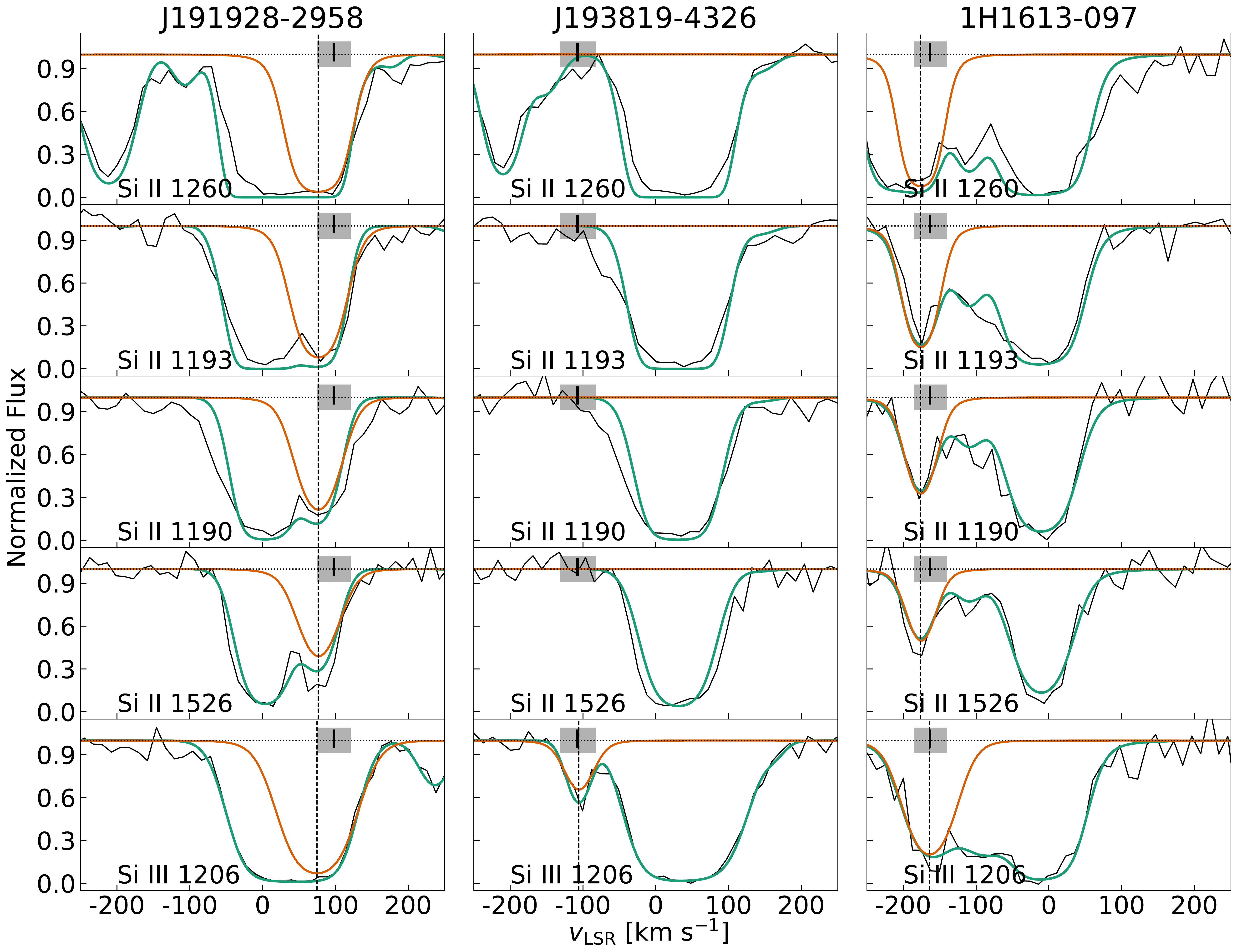}
    \end{center}
\caption{Profiles of the \SiII\ and \SiIII\ absorption lines 
for J1919-2958, J1938-4326, and 1H1613-098. 
The black tick marks and grey bars at the top of every panel represents the velocity centroid and three times its error of the \ion{O}{i} (we include a COS zero-velocity offset in the v$_{0\:\mathrm{UV}}$ for 1H1613-097, which is not reported in the literature table from which the velocity centroid was drawn\cite{Bordoloi_2017}). 
The Voigt-profile fits 
to the \SiII\ and \SiIII\ absorption are denoted by the orange lines when detected and the full fit to each absorption spectra is given by the green line. The velocity centroids of the \SiII\ and \SiIII\ fits are marked with a dashed vertical line.}\label{figure:si_ions}
\end{figure*}

\clearpage
\section*{Supplementary information references}\vspace{12pt}
\begin{enumerate}
  \setcounter{enumi}{67}

\item{Sofia, U.~J., Parvathi, V.~S., Babu, B. R.~S. \& Murthy, J. Determining interstellar carbon abundances from strong-line transitions. \emph{\aj} \textbf{141}, 22, (2011).}

\end{enumerate}

%TC:endignore

\end{document}